\documentclass[twocolumn,prb,showpacs,floatfix,superscriptaddress]{revtex4}
\usepackage{graphicx}
\usepackage{bm}
\usepackage{amsmath,amssymb}
\usepackage{hyperref}

\newcommand{\be}{\begin{equation}}
\newcommand{\ee}{\end{equation}}
\newcommand{\bea}{\begin{eqnarray}}
\newcommand{\eea}{\end{eqnarray}}

\begin{document}

\title{Classical and quantum dimers on the star lattice}

\author{John Ove Fj{\ae}restad}
\affiliation{Department of Physics, The University of Queensland,
Brisbane, QLD 4072, Australia}

\date{November 24, 2008}

\begin{abstract}

We consider dimers on the star lattice (aka the 3-12, Fisher, expanded kagome or triangle-honeycomb lattice).
We show that dimer coverings on this lattice have $Z_2$ arrow and pseudo-spin representations analogous to
those for the kagome lattice, and use these to construct an exactly solvable quantum dimer model (QDM) with a
Rokhsar-Kivelson (RK) ground state. This QDM, first discussed by Moessner and Sondhi from a different point of view,
is the star-lattice analogue of a kagome-lattice QDM analyzed by Misguich \textit{et al}. We give a detailed
analysis of various properties of the classical equal-weight dimer model on the star lattice, most of which are
related to those of the RK state. Using both the arrow representation and the fermionic path integral formulation
of the Pfaffian method, we calculate the number of dimer coverings, dimer occupation
probabilities, and dimer, vison, and monomer correlation functions. We show that a dimer is uncorrelated with dimers
further away than on neighboring triangles, that the vison-vison correlation function vanishes, and that the
monomer-monomer correlation function equals $1/4$ regardless of the monomer positions (with one exception).
A key result in the fermionic approach is the vanishing of the two-point Green function beyond a short distance.
These properties are very similar to those of dimers on the kagome lattice. We also discuss some generalizations
involving arrow representations for dimer coverings on ``general Fisher lattices'' and their ``reduced'' lattices,
with the kagome, squagome, and triangular kagome lattice being examples of the latter.

\end{abstract}


\maketitle

\section{Introduction}
\label{intro}

\begin{figure}[htb]
\begin{center}
\centerline{\includegraphics[scale=0.7]{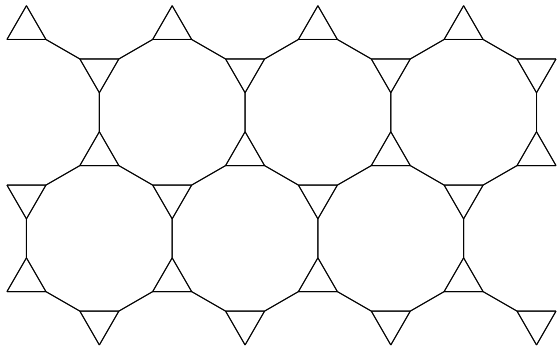}}
\caption{The star lattice. The lattice sites are the vertices in this figure.
The lines connecting the sites will be called bonds. There are two types
of bonds, which will be referred to as triangle bonds and linking bonds. A triangle
bond is part of a triangle. A linking bond connects two neighboring triangles.}
\label{fig:star-lattice}
\end{center}
\end{figure}

Classical dimer models (CDMs) have long been of interest in statistical and
theoretical physics.\cite{k-67,wu-review,fms-02,dijkgraaf,math} An important
breakthrough was ushered in with the solution of the dimer problem on the square
lattice in terms of Pfaffians.\cite{k-61,tf-61,fs-63} In a subsequent seminal
paper,\cite{k-63} Kasteleyn generalized this solution method to the dimer problem on any planar lattice,
and also showed explicitly that CDMs can exhibit singularities, corresponding to phase transitions,
as a function of the dimer weights. A few years later Fisher showed that the Ising
model on an arbitrary planar lattice can be mapped to a CDM on a different planar lattice,\cite{f-66}
thus making it possible to solve the former class of models with the Pfaffian method developed
for the latter. More recently, Moessner and Sondhi\cite{ms-03} have demonstrated that the nature
of the phases and phase transitions in the two models connected by this mapping can be very different.
Specifically, they revisited Fisher's mapping between the ferromagnetic Ising model on the square lattice
and the CDM on the star lattice\cite{name} (shown in Fig. \ref{fig:star-lattice}) and found that the
symmetry-breaking transition in the former model maps to a confinement-deconfinement transition
for test monomers in the latter.

Over the last two decades, quantum dimer models (QDMs) have come to
play an important role in the search for phases with exotic
types of orders and excitations in quantum antiferromagnets (for recent
reviews, see Refs. \onlinecite{r-et-al-rev} and \onlinecite{mr-rev}).
Most intriguingly, there is the possibility of spin liquid phases\cite{and-faz}
with deconfined, fractionalized (spin-$1/2$ \textit{spinon}) excitations. These spin
liquids do not break spin rotation or spatial symmetries but are instead characterized
by unconventional \textit{quantum orders}\cite{wen-book} (e.g. topological order) not
describable by local order parameters. The canonical example of a spin liquid is a short-range
resonating-valence-bond (RVB) state\cite{and-faz} given by a macroscopic superposition
of products of spin singlets formed by pairs of nearest-neighbor spins.
QDMs were introduced by Rokhsar and Kivelson\cite{rk} to be an approximate,
effective description of the low-energy physics when it is dominated by such
short-range singlet fluctuations.

The set of dimer coverings in a CDM is taken to form an orthogonal basis for the
Hilbert space of QDMs on the same lattice. A QDM Hamiltonian can contain both potential
and kinetic energy terms, which are respectively diagonal and off-diagonal in this
dimer basis. If these terms have amplitudes $v$ and $t$ respectively, the QDM ground state
will be a function of $v/t$. This parameter space generally includes a so-called
Rokhsar-Kivelson (RK) point, at which the ground state is an equal-amplitude
superposition of dimer coverings (in each topological sector).\cite{rk} As a consequence,
there is a close relationship between QDMs at the RK point and CDMs:
any ground state correlation function at the RK point that only involves operators
which are diagonal in the dimer basis (e.g., the dimer-dimer or vison-vison
correlation function) is equal to the corresponding correlation function
for the CDM with equal dimer weights on the same lattice.\cite{rk}

The search for dimer liquids in QDMs was crowned with success when Moessner and
Sondhi\cite{ms-rvb} showed that the ground state of the triangular-lattice QDM is a dimer
liquid (i.e. has exponentially decaying dimer-dimer correlations) over a finite range of
$v/t$. In another remarkable paper, Misguich \textit{et al.}\cite{mis-kag-1} introduced and
analyzed a QDM on the kagome lattice that differed in several respects from the QDMs
studied until then. This kagome QDM does not have potential energy terms, and
its kinetic energy involves flippable loops\cite{flippable} of various lengths
(in contrast, the prototypical QDM studied in Refs. \onlinecite{rk} and \onlinecite{ms-rvb}
only includes the shortest flippable loops). Due to special properties of the kagome
lattice, which allow for $Z_2$ arrow and pseudo-spin representations of its dimer
coverings,\cite{ez-93,mis-kag-1,mis-kag-2}
Misguich \textit{et al.}\cite{mis-kag-1} were able to show that this QDM can be written in an
extremely simple form in terms of the pseudo-spin variables. As a consequence, all its eigenstates and
eigenvalues are exactly known. In particular, its ground state in each topological sector is the
RK (equal-amplitude) state. On the kagome lattice this state is an extremely disordered dimer liquid:
the dimer-dimer correlations vanish for dimers further apart than on neighboring
triangles. This has  been shown from the arrow representation\cite{mis-kag-1} as
well as from Pfaffian calculations of the dimer-dimer correlations of the CDM on the
kagome lattice.\cite{wang-wu-07,wang-wu-08} Another interesting result for the CDM on
the kagome lattice is related to monomer deconfinement: the monomer-monomer correlation
function equals $1/4$ regardless of the distance between the two monomers. Again, this has been shown
both from the arrow representation\cite{mis-lhu} and from Pfaffian calculations.\cite{lyc-08}

In this paper we consider dimers on the star lattice\cite{name} (Fig. \ref{fig:star-lattice}), a non-bipartite
lattice that is one of the 11 Archimedean tilings.\cite{arch} Because of its frustration\cite{richter-review}
and small coordination number ($z=3$) this is an interesting lattice to consider in the search for unconventional
phases in spin and dimer models. Various studies of this kind have recently appeared in the literature.
We have already mentioned
Ref. \onlinecite{ms-03} which we will come back to shortly. From an analysis of exact diagonalization
spectra Richter \textit{et al.}\cite{richter-review} concluded that the kagome and star lattices are the
only Archimedean tilings for which the ground state of the nearest-neighbor spin-$1/2$ Heisenberg model does
not have magnetic long-range order. For the star lattice the ground state is instead a unique valence bond
crystal (VBC) with singlets across the bonds that link different triangles.\cite{richter-review,richter-2}
If the exchange across these linking bonds is made sufficiently small compared to that of the triangle bonds,
exact diagonalization results suggest that the ground state is a different, three-fold degenerate VBC.\cite{mis-sin}
We also note that an experimental realization of a star antiferromagnet,
[Fe$_3$($\mu_3$-O)($\mu$-OAc)$_6$-(H$_2$O)$_3$][Fe$_3$($\mu_3$-O)($\mu$-OAc)$_{7.5}$]$_2$ $\cdot 7\,$H$_2$O,
has been reported recently.\cite{star-exp} Furthermore, Yao and Kivelson\cite{yao-kivelson} have
shown that the ground state of the Kitaev model on the star lattice is a chiral spin liquid.

We show that dimers on the star lattice have properties that are very similar to those for
the kagome lattice summarized above. We first show that
dimer coverings on the star lattice have arrow and pseudo-spin representations analogous to those
for the kagome lattice. We use these to construct a star-lattice QDM that has similar properties as
the kagome QDM discussed by Misguich \textit{et al.}\cite{mis-kag-1} In particular, its ground state
is the RK state. This star-lattice QDM was first discussed in Ref. \onlinecite{ms-03} as a special
case of a QDM introduced there via a modified Fisher construction mapping it to a three-dimensional
Ising model in the time-continuum limit.
We give a detailed analysis of the star-lattice CDM for equal-weight dimers (the case related to the
RK ground state as mentioned above), using both the Pfaffian method and the arrow representation to calculate
the number of dimer coverings (on a finite lattice without boundaries), dimer occupation probabilities,
and dimer, vison, and monomer correlation functions.

For our Pfaffian calculations we use Samuel's formulation based on a path integral description of free
Majorana fermions.\cite{samuel-1} We find that the (two-point) Green function of the fermions vanishes
beyond a short distance. This is a key result that causes a dimer to be uncorrelated with all dimers further
away than on a neighboring triangle. It furthermore makes the calculations of the vison-vison and
monomer-monomer correlation functions tractable; although in the fermionic formulation these functions
contain a ``string'' going from one vison/monomer to the other, which makes the number of pairings from Wick's
theorem grow exponentially with the vison/monomer separation, the number of \textit{nonzero} pairings is strongly
reduced by the above property of the Green function. We also use the arrow representation to calculate the
various quantities considered. These arrow derivations are conceptually simpler and more intuitive than the
Pfaffian ones, in part because they lend themselves to a pictorial representation. Finally, we consider some
generalizations to general Fisher lattices (any lattice hosting the dimers in Fisher's mapping\cite{f-66}),
and discuss ``reduced'' lattices of these (examples of which include the kagome, squagome, and
triangular kagome lattice).

The paper is organized as follows. The arrow and pseudo-spin representation
are introduced in Secs. \ref{arrow} and \ref{ps}. In Sec. \ref{ps} we
also construct the exactly solvable QDM. The number of dimer coverings is considered
in Sec. \ref{ndc}. The Green function is calculated in Sec. \ref{green}. In Secs.
\ref{d-occ} and \ref{d-corr} we discuss dimer occupation probabilities and dimer correlations.
The vison-vison and monomer-monomer correlation functions are calculated in Secs. \ref{vv-corr}
and \ref{mm-corr}. In Sec. \ref{comp-kagome} we summarize similarities and connections
between dimers on the star and kagome lattice. In Sec. \ref{gen-fisher} we discuss properties of
general Fisher lattices and their reduced lattices (with some technical details relegated to
Appendix \ref{app-fisher}). Some concluding remarks are given in Sec. \ref{conc}. In
Appendix \ref{dc-gen} we present an alternative pseudo-spin representation of dimer coverings
on the star lattice.

\section{Arrow representation of dimer coverings}
\label{arrow}

In this section we show that there is a one-to-one correspondence between dimer
coverings and arrow configurations on the star lattice. The arrows are Ising
degrees of freedom located on the lattice sites. First, however, we establish
some basic terminology that will be used throughout the paper.

The star lattice consists of triangles and dodecagons as shown in Fig. \ref{fig:star-lattice}.
The lattice sites sit at the vertices of the polygons. The thin lines connecting different sites will
be referred to as bonds. While all sites in the star lattice have the same local environment (by
definition of an Archimedean tiling), there are two types of inequivalent
bonds, which will be referred to as triangle bonds and linking bonds.
A triangle bond is part of a triangle (as well as a dodecagon) while a
linking bond is part of two dodecagons and is a link between two triangles.

\begin{figure}[htb]
\begin{center}
\centerline{\includegraphics[scale=0.9]{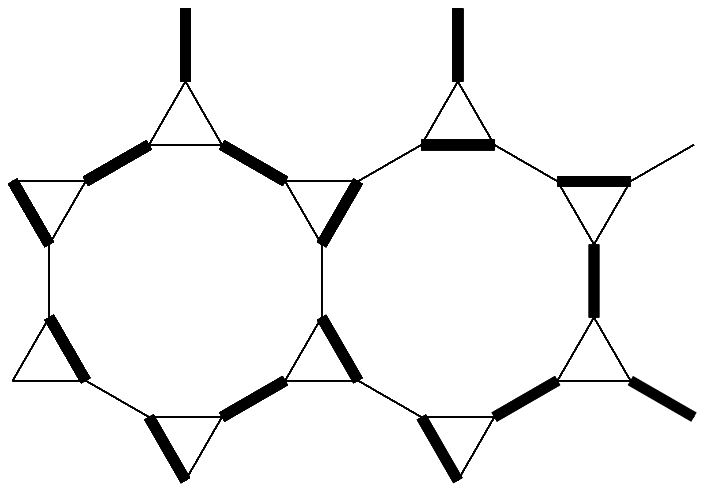}}\vspace{0.4cm}
\centerline{\includegraphics[scale=0.9]{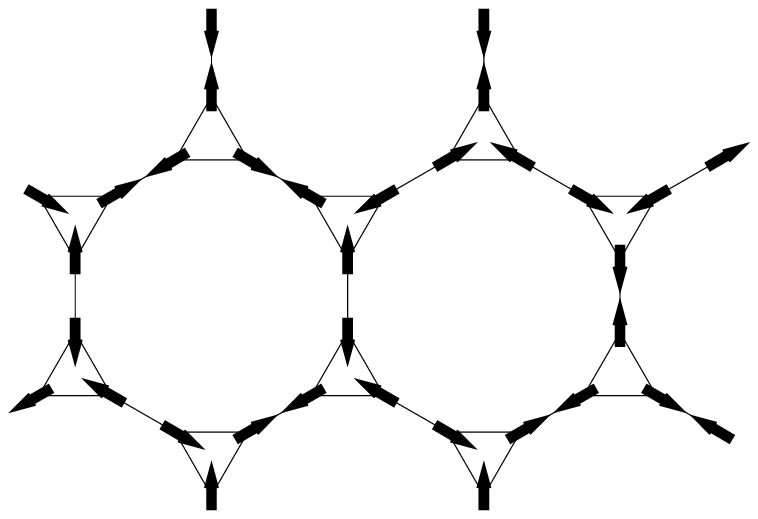}}
\caption{Top: Example of dimer covering on the star lattice. Dimers are shown as
thick lines. Bottom: The same dimer covering in the arrow representation.}
\label{fig:star-dimer}
\end{center}
\end{figure}

A bond can be occupied by a dimer, which will be drawn as a thicker line on the
bond. The dimer touches the two sites that are connected by the bond. We will
consider close-packed, hard-core dimer configurations. Close-packed means
that no site is left untouched by a dimer. Hard-core means that no site is
touched by more than one dimer. The fact that each site is touched by
exactly one dimer will be referred to as the closed-packed, hard-core constraint.
In the following we will refer to such dimer configurations simply as
dimer coverings. (In Sec. \ref{mm-corr} we will relax the
close-packed part of the constraint to consider dimer-monomer configurations in which two sites
in the lattice are not touched by a dimer.) An example of a dimer covering
on the star lattice is shown in Fig. \ref{fig:star-dimer}(top).

Elser and Zeng showed that dimer coverings on the kagome
lattice allow for an arrow representation.\cite{ez-93} Misguich \textit{et
al.} made extensive use of this representation in their work on QDMs on the
kagome lattice,\cite{mis-kag-1,mis-kag-2} and also
pointed out that the arrow representation can be generalized further to all
lattices made of corner-sharing triangles. We will now show that an arrow
representation exists for dimer coverings on the star lattice too
(arrow representations for some other lattices will be discussed in
Sec. \ref{gen-fisher} and \ref{conc}).

In the arrow representation for the star lattice, an arrow is located on
each lattice site. It can point in one of two directions: either towards
the center of the triangle to which the site belongs, or in the opposite
direction, towards the center of the site's linking bond. In the
first case the dimer that touches the site lies on one of the site's
two triangle bonds. In the second case the dimer lies on the site's
linking bond. If a triangle contains a dimer on bond $ij$ the
arrows on sites $i$ and $j$ both point into the triangle while the
arrow on the triangle's third site points out of the triangle. In
contrast, if a triangle has no dimer, all three arrows point out of
the triangle. As these are the only possibilities, each triangle
must satisfy the constraint that an even number (0 or 2) of its
arrows must point into the triangle. Next consider a linking bond
$kl$. If it has a dimer, the arrows on sites $k$ and $l$ both point
"into" (i.e. towards the center of) the linking bond. Alternatively,
the linking bond has no dimer, in which case the directions of both
arrows are reversed to point "out of" (i.e. away from the center of)
the linking bond. Hence each linking bond must satisfy the
constraint that an even number (0 or 2) of its arrows must point
"into" the bond. The constraint on a triangle is exactly
the same as for the lattices of corner-sharing triangles for which
the arrow representation has been used earlier. Furthermore, the
new constraint introduced for linking bonds takes the same form as the triangle
constraint: an even number of arrows must point inwards. An example
of a dimer covering and its arrow representation is given in Fig.
\ref{fig:star-dimer}.

\section{Pseudo-spin representation of dimer coverings and an
exactly solvable quantum dimer model}
\label{ps}

Elser and Zeng introduced an Ising pseudo-spin representation for dimer
coverings on the kagome lattice.\cite{ez-93} Misguich
\textit{et al.}\cite{mis-kag-1,mis-kag-2} developed this representation
further by unveiling additional important properties. In this section
we show that an analogous pseudo-spin representation exists for the star
lattice.\cite{mis-sin-men} In Secs. \ref{ps-z} and \ref{ps-x} we derive
the properties of this representation. In Sec. \ref{exact-qdm} we use the pseudo-spin representation
to define a star lattice QDM whose eigenstates and energy spectrum are
exactly known and which has a Rokhsar-Kivelson ground state. This
model has previously been discussed by Moessner and Sondhi from
a different point of view.\cite{ms-03}

\subsection{The operators $\hat{\sigma}^z(D)$}
\label{ps-z}

We consider a star lattice on a closed surface, i.e. a system
without boundaries. The dimer coverings (DCs) can be grouped into
topological sectors. To do this one looks at the transition graph
of two DCs obtained by superposing them on top of each other. These
transition graphs may contain two different types of non-intersecting
even-length closed loops: loops that enclose a nonzero area and
loops that enclose zero area (loops of the latter type have length 2).
Loops of the first type can be topologically trivial or nontrivial
(all loops of the second type are obviously topologically trivial).\cite{top-loop}
If the transition graph of two DCs has no topologically non-trivial
loops, these DCs are in the same topological sector.

Next consider a given topological sector for DCs on the star lattice.
Let $|c_0\rangle$ be some arbitrary but specific DC in this topological sector; it will play
the role of a reference DC. Introduce a spin-$1/2$ operator $\hat{\sigma}^z(D)$
for each dodecagon $D$ with eigenvalues $\sigma^z(D)=\pm 1$. To any DC
$|c\rangle$ in the same topological sector as $|c_0\rangle$ we can associate a
pair of pseudo-spin configurations (PSCs) $\{\sigma^z(D)\}$ and $\{-\sigma^z(D)\}$ related
by a global spin reversal, as follows: Draw the transition graph $\langle c_0|c\rangle$.
Consider the closed loops that enclose a nonzero area in this graph; these loops
are taken to be domain walls separating dodecagons with opposite values of $\sigma^z$.\cite{consistent}
The need to identify the two PSCs related by a global spin reversal with the same DC is due to the fact
that the domain walls only determine the value of the \textit{products} $\sigma^z(D)\sigma^z(D')$
for all dodecagons $D$, $D'$, which are invariant under a global spin reversal. The
reference DC $|c_0\rangle$ is associated with the pair of PSCs having all $\sigma^z$ up and
all $\sigma^z$ down.

The mapping from DCs in a given topological sector to pairs of PSCs can also
be formulated in terms of the arrow representation. Let $D$ and $D'$ be two
neighboring dodecagons and consider the linking bond separating $D$ and $D'$.
If the two arrows associated with this linking bond point in opposite (the same)
directions in $|c\rangle$ and $|c_0\rangle$, there will (will not) be a domain
wall between $D$ and $D'$ in the transition graph $\langle c_0|c\rangle$, so
that the product $\sigma^z(D)\sigma^z(D')=-1$ ($1$) in the pair of PSCs associated
with $|c\rangle$.

Two different DCs $|c\rangle$ and $|c'\rangle$ in the same topological sector
will  map to different pairs of PSCs. To see this we note that if the DCs are different, there
will be some linking bonds that are occupied by a dimer in $|c\rangle$ but unoccupied
in $|c'\rangle$ and vice versa. (In other words, it is impossible to have dimer
coverings that differ only in the dimer occupation numbers on triangle bonds).
Thus the arrows on these linking bonds point in opposite directions in $|c\rangle$
and $|c'\rangle$, and therefore the product $\sigma^z(D)\sigma^z(D')$ for the
dodecagons $D$ and $D'$ separated by each such linking bond will be different for
the PSCs of $|c\rangle$ and $|c'\rangle$.

Any pair of PSCs related by global spin reversal corresponds to a unique
DC $|c\rangle$ in the topological sector of $|c_0\rangle$ (this DC will however depend on the
choice of $|c_0\rangle$). To see this we note that given these PSCs we can find the
value of the product $\sigma^z(D)\sigma^z(D')$ for each neighboring pair of dodecagons
$D$ and $D'$. If the product is $-1$ ($1$) the directions of the two arrows on the linking bond
separating $D$ and $D'$ are opposite (the same) in $|c\rangle$ and $|c_0\rangle$.
Since every arrow is associated with a unique linking bond in this way, all arrows are
assigned a direction by this procedure, thus uniquely specifying $|c\rangle$.

From this mapping between DCs in a given topological sector and pairs of PSCs
$\{\sigma^z(D)\}$ and $\{-\sigma^z(D)\}$, it follows
that the number of DCs in a topological sector of the star lattice is the
same for each sector and given by $2^{N_D-1}$ where $N_D$ is the number of dodecagons. This
result can be expressed in terms of the number of sites $N$ and the genus $g$ of the manifold by using Euler's formula
$V+F-E=\chi$ where $V$, $F$, and $E$ are the number of vertices, faces, and edges in the
lattice graph and the Euler characteristic $\chi=2-2g$ for a closed orientable surface
of genus $g$. Here $V=N$, $F=N_D+N_T$ where $N_T=N/3$ is the number of triangles, and
$E=N_t+N_l$ where $N_t=3N_T=N$ is the number of triangle bonds and $N_l=N/2$ is the
number of linking bonds. This gives $N_D = N/6+2-2g$ so the number of DCs in each
topological sector is $2^{N/6+1-2g}$. In Sec. \ref{ndc-arrows} we will show that the
total number of DCs on this lattice graph (i.e. including all topological sectors) is
$2^{N/6+1}$. Thus the number of topological sectors is $2^{N/6+1}/2^{N/6+1-2g}=2^{2g}=4^g$,
which is the expected result for a non-bipartite lattice on a genus-$g$ manifold.

\subsection{The operators $\hat{\sigma}^x(D)$}
\label{ps-x}

Using the arrow constraints, one finds that there are
64 different configurations for the 18 arrows on the six triangles that surround a given
dodecagon. Each of these arrow configurations corresponds, in the dimer picture, to
a ``flippable''\cite{flippable} dimer configuration along one of the 32 even-length loops that enclose
the dodecagon and an even number of its six surrounding triangles (for each such loop there are two
flippable dimer configurations). There is 1 loop of length 12, 15 loops of length 14, 15 loops
of length 16, and 1 loop of length 18. The loops in these four classes surround the dodecagon and 0,
2, 4, and 6 of its neighboring triangles, respectively.

We now define the operator $\hat{\sigma}^x(D)$ as
\be
\hat{\sigma}^x(D) = \sum_{\alpha}(|L_{\alpha}(D)\rangle \langle \bar{L}_{\alpha}(D)|+\mbox{h.c.}).
\label{sigma-x}
\ee
This definition is analogous to the one for the kagome lattice.\cite{mis-kag-1,mis-kag-2}
The sum runs over the 32 even-length loops associated with $D$. $|L_{\alpha}(D)\rangle$ and
$|\bar{L}_{\alpha}(D)\rangle$ are the two ``flippable'' dimer configurations along loop $\alpha$ around dodecagon $D$.
It can be seen that, in the arrow representation,
$\hat{\sigma}^x(D)$ flips the direction of the 12 arrows sitting on the dodecagon $D$. This operation
respects the arrow constraints since it flips an even number (0 or 2) of the arrows involved in each arrow
constraint in the system. From this result it immediately follows that $(\hat{\sigma}^x(D))^2=I$ (flipping the arrows
twice is equivalent to doing nothing) and that $[\hat{\sigma}^x(D),\hat{\sigma}^x(D')]=0$
(it doesn't matter in which order the arrows are flipped).

Next let $D'$ be any of the dodecagons neighboring $D$. Consider the effect of acting with $\hat{\sigma}^x(D)$
on a DC $|c\rangle$ in the same topological sector as $|c_0\rangle$. If the directions of the two arrows on the
linking bond separating $D$ and $D'$ are opposite (the same) in $|c\rangle$ and $|c_0\rangle$, they will be the same
(opposite) in $\hat{\sigma}^x(D)|c\rangle$ and $|c_0\rangle$. Thus for all $D'$ that neighbor $D$,
$\hat{\sigma}^x(D)$ anticommutes with $\hat{\sigma}^z(D)\hat{\sigma}^z(D')$. Furthermore, if $D_a$ and $D_b$
are two neighboring dodecagons, both different from $D$, then $\hat{\sigma}^z(D_a)\hat{\sigma}^z(D_b)$ clearly
commutes with $\hat{\sigma}^x(D)$ since the latter doesn't affect the arrows on the linking bond separating
$D_a$ and $D_b$. These two results can be generalized\cite{gen-comm} to, respectively, $D'$ not a neighbor of $D$ and
$D_a$ and $D_b$ not neighbors of each other. From this we conclude that $\hat{\sigma}^x(D)$ anticommutes with
$\hat{\sigma}^z(D)$ and commutes with $\hat{\sigma}^z(D')$ for $D'\neq D$. This shows that (as implied by the notation)
$\hat{\sigma}^x(D)$ is the spin-$1/2$ operator that flips the pseudo-spin $\sigma^z(D)$ at dodecagon $D$.
Also note that
\be
\prod_{D}\hat{\sigma}^x(D)=I.
\label{sigmax-constraint}
\ee
This constraint involving all dodecagons follows from the fact that $\prod_D \hat{\sigma}^x(D)$ is the
operator that effects a global spin reversal, under which a PSC $\{\sigma^z(D)\}$ maps to its partner
$\{-\sigma^z(D)\}$ that represents the same DC.
Alternatively, this constraint can be understood from the arrow representation, since
the lhs of Eq. (\ref{sigmax-constraint}) flips all arrows twice and therefore is equivalent to doing nothing.

Starting from the arrow representation for the reference DC $|c_0\rangle$, and given one of
the two PSCs for the DC $|c\rangle$ in the same sector, the arrow representation for $|c\rangle$ can be
obtained either by applying $\hat{\sigma}^x$ to $|c_0\rangle$ at all dodecagons with $\sigma^z=+1$ in $|c\rangle$ or
by applying $\hat{\sigma}^x$ to $|c_0\rangle$ at all dodecagons with $\sigma^z=-1$ in $|c\rangle$. Either way,
the arrows that point in different directions in $|c_0\rangle$ and $|c\rangle$ will be flipped once, while the
arrows that point in the same direction in $|c_0\rangle$ and $|c\rangle$ will be flipped zero or two times.

\subsection{An exactly solvable quantum dimer model}
\label{exact-qdm}

Given the form of $\hat{\sigma}^x(D)$, it is natural to consider a QDM defined by the Hamiltonian
\be
\hat{H}=-\Gamma\sum_D \hat{\sigma}^x(D)
\label{qdm-ham}
\ee
with $\Gamma$ a positive constant. This QDM, which has no potential energy terms, is the star-lattice analogue
of the kagome lattice QDM introduced by Misguich \textit{et al.}\cite{mis-kag-1} It has been discussed previously
by Moessner and Sondhi\cite{ms-03} as a particular limit of a QDM they arrived at via a mapping (based on a modified
Fisher construction) to a three-dimensional Ising model in the time-continuum limit.

Let us discuss some basic properties of the QDM (\ref{qdm-ham}) from the point of view of the pseudo-spin
representation that we have introduced in this section. Clearly the eigenstates are
of the form $\prod_D |\sigma^x(D)\rangle$ where $|\sigma^x(D)\rangle$ is an eigenstate of $\hat{\sigma}^x(D)$
with eigenvalue $\sigma^x(D)=\pm 1$. Note however that due to the constraint (\ref{sigmax-constraint}),
only eigenstates with an \textit{even} number of dodecagons in the excited $\sigma^x=-1$ state are allowed.
These $\sigma^x=-1$ excitations localized at the dodecagons are (point) visons.\cite{mis-kag-1,ms-03}
We will discuss them further in Sec. \ref{vv-corr}. In the ground state, each dodecagon is in the $\sigma^x=+1$
state which is just the
sum of the $\sigma^z=+1$ and $\sigma^z=-1$ eigenstates. Expanding out the product over dodecagons, the
ground state is seen to be the equal-amplitude superposition of all dimer coverings in a given topological sector,
i.e. the RK state. Since the ground state energy is the same in each of the $4^g$ topological sectors the system has
topological order.\cite{top-order,wen-book} (Note that the ground state is degenerate even for a system with a
finite number of sites.)

The fact that the ground state is $4^g$-fold degenerate can also be seen as follows:
A ground state $|\Psi_0\rangle$ of (\ref{qdm-ham}) satisfies $\hat{\sigma}^x(D)|\Psi_0\rangle =|\Psi_0\rangle$ for each
dodecagon $D$. This can be regarded as a constraint on $|\Psi_0\rangle$. There is one such
constraint for each dodecagon. However, due to (\ref{sigmax-constraint}) only $N_D-1$ of these constraints
are independent. Every independent dodecagon constraint reduces the number of possible states in the ground
state manifold by a factor of two, since it specifies one of two possible eigenstates at that dodecagon.
As in Sec. \ref{ps-z} we invoke the result to be shown in Sec. \ref{ndc-arrows}, that the total number of dimer
coverings is $2^{N/6+1}$ which is therefore also the dimension of the Hilbert space. Thus the dimension of the
ground state manifold is $2^{N/6+1-(N_d-1)}=4^g$, where we used $N_D=N/6+2-2g$ as shown in Sec. \ref{ps-z}.

As noted in the introduction, RK-state correlations involving operators that are diagonal in the dimer basis
are given by the corresponding correlation functions in the CDM with equal dimer weights. This CDM will be
analyzed in the following sections.

\section{The number of dimer coverings}
\label{ndc}

In this section we consider the number of dimer coverings $\mathcal{Z}$
on a star lattice graph with a finite number of sites $N$. We will
take the graph to be embedded on a closed surface so that the system
has no boundaries. We first consider a surface of arbitrary genus
$g$ and give a derivation of $\mathcal{Z}$ based on the arrow representation
introduced in Sec. \ref{arrow}. Next we calculate $\mathcal{Z}$ for a torus
(genus $g=1$) using the Pfaffian method. Some alternative derivations of
this result are discussed in Sec. \ref{gen-fisher} and Appendix \ref{dc-gen}.

\subsection{Arrow derivation for a genus-$g$ manifold}
\label{ndc-arrows}

Consider a star lattice graph embedded on a closed orientable surface
of genus $g$. If this graph has $N$ sites there are $N/3$ triangles and
$N/2$ linking bonds. As the two latter numbers must both be integers, $N$ is a multiple
of $6$ (and thus also even). Since there is one arrow per site and each arrow is an Ising
variable (i.e. it can point in two different directions) it follows
that in the absence of any constraints there would be $2^{N}$
different ways to choose the directions of the arrows. Each of the
triangles and linking bonds does however contribute an arrow constraint,
giving a total of $N/3+N/2=5N/6$ constraints. As will be shown below,
one of these constraints can be deduced from the others, leaving
$5N/6-1$ independent constraints, each of which can be used to eliminate
exactly one arrow variable.\cite{eliminate} Thus the total number of
dimer coverings on this star lattice graph is
\be
\mathcal{Z}=2^{N-(5N/6-1)}=2^{N/6+1}.
\label{Zs}
\ee
We note that the result is independent of the genus $g$. The associated
entropy $\mathcal{S}=\log \mathcal{Z}$ is thus
\be
\mathcal{S}=\frac{\log 2}{6}N + \log 2.
\label{ent-1}
\ee
As shown in Sec. \ref{ps-z} there are $4^g$ topological sectors, each of
which have the same number of dimer coverings $\mathcal{Z}_{\rm{ts}} = \mathcal{Z}/4^{g}$.
The corresponding entropy per sector is therefore
\be
\mathcal{S}_{\rm{ts}}=\frac{\log 2}{6}N + (1-2g)\log 2.
\label{ent-2}
\ee
From these expressions one sees that the entropy per site in the thermodynamic limit is given
by $(1/6)\log 2$, in agreement with previous calculations.\cite{wu-review,per-dimer} In addition to the
leading $O(N)$ (i.e., extensive) term the entropy also has a sub-leading $O(1)$ term that carries information
about the topology of the system.

Let us show that only $5N/6-1$ arrow constraints are independent as
claimed above. Consider an arbitrary site $i$. Denote the triangle
and the linking bond it belongs to by $T$ and $l$ respectively. Define
the arrow variable $a_{i,T}=+1$ $(-1)$ if the arrow on site $i$
points out of (into) $T$. Similarly, define the arrow variable
$a_{i,l}=+1$ ($-1$) if the arrow points out of (into) $l$. Clearly
these two variables are related by $a_{i,T}=-a_{i,l}$. Furthermore
define $a_T = \prod_{i\in T}a_{i,T}$ where the product runs over the
three sites belonging to the triangle $T$, and $a_l\equiv
\prod_{i\in l}a_{i,l}$ where the product runs over the two sites
belonging to the linking bond $l$. We have
\be
\left(\prod_T a_T\right)\left(\prod_l
a_l\right)=\prod_{i=1}^{N}a_{i,T}a_{i,l}=(-1)^{N}=1.
\label{arrow-prodall}
\ee
On the other hand, the $5N/6$ arrow constraints can be written
\be
a_T = a_l = 1
\mbox{ for all }T,\;l. \label{arrow-con}
\ee
From (\ref{arrow-prodall}) we see that one of the
constraints in (\ref{arrow-con}) follows from the others,
leaving $5N/6-1$ independent constraints.

\subsection{Pfaffian calculation for a torus}
\label{pfaff-torus}

The Pfaffian method can be used to calculate the number of dimer
coverings $\mathcal{Z}$ on any planar lattice graph.\cite{k-63}
In this method the central object is an antisymmetric
matrix $A$, sometimes called a Kasteleyn matrix, whose dimension
equals the (even) number of sites $N$. The matrix
element $A_{ij}$ is nonzero only if sites $i$ and $j$ are connected
by a bond in the lattice graph, in which case the magnitude of $A_{ij}$
is set equal to 1.\cite{general} The sign of $A_{ij}$
defines the direction of an arrow\cite{arrow-confuse} on the bond $ij$:
If $A_{ij}>0$ the arrow points from $i$ to $j$. The signs of the matrix elements
of $A$ should be chosen to satisfy Kasteleyn's clockwise-odd rule:\cite{k-63,detailed}
The perimeter of each face of the lattice graph should contain an odd
number of arrows pointing in the clockwise direction
around the face. One then has $\mathcal{Z}=|\mbox{Pf }A|$,
where $\mbox{Pf }A$ is the Pfaffian of $A$. Using $(\mbox{Pf }A)^2=\det A$
it follows that $\mathcal{Z}=\sqrt{\det A}$ for a planar lattice graph.

\begin{figure}[!htb]
\begin{center}
\includegraphics[scale=0.75]{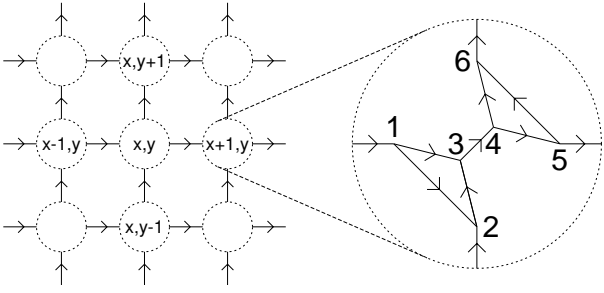}
\caption{The star lattice drawn as a square lattice (left) of
6-site unit cells (right). Also shown are the directions of
the Kasteleyn arrows and the labeling of unit cells and sites
within a unit cell. These conventions follow Ref. \onlinecite{f-66}.}
\label{fig:star-fisher}
\end{center}
\end{figure}

Fisher\cite{f-66} and Wu\cite{wu-review} have used the Pfaffian method to calculate
the free energy per site/dimer for the dimer problem on an \textit{infinite} star lattice. For our
Pfaffian calculations we follow Fisher in viewing the star lattice as a square lattice
of 6-site unit cells, as shown in Fig. \ref{fig:star-fisher}, where we also use his labeling of
sites and choice of bond arrow directions to satisfy Kasteleyn's sign rule. A site $i$
is written $i=\bm{r},\alpha$ where $\alpha=1,\ldots,6$ is the site label within the unit cell,
and $\bm{r}=(x,y)$ labels the unit cell, where $x=1,\ldots,N_x$ is the horizontal coordinate and
$y=1,\ldots,N_y$ the vertical one, with $N_x$ and $N_y$ being the number of unit cells in the
horizontal and vertical direction, respectively. The total number of sites in the lattice is
then $N=6N_x N_y$, which is always even so that the lattice can be completely covered by dimers
for any choice of $N_x$ and $N_y$.

When the sites on opposite edges in the two directions are connected by
additional bonds to form a torus, the resulting lattice graph is no longer planar.
However, as shown by Kasteleyn,\cite{k-61} the dimer problem on a torus can
still be solved using Pfaffians by a small modification of
the procedure described so far. In this case $\mathcal{Z}$
can be expressed as a linear combination of
the Pfaffians of four different Kasteleyn matrices $A^{\nu_x\nu_y}$ ($\nu_x,\nu_y=0,1$),
where the first (second) superscript on $A^{\nu_x\nu_y}$ labels the type of
boundary condition imposed on the matrix elements in the horizontal (vertical) direction.
Compared to the single Kasteleyn matrix for the open-boundary case, these
four matrices differ only in the presence of additional nonzero matrix elements corresponding
to the additional bonds introduced to establish the torus geometry. Specifically, the
new matrix elements of $A^{\nu_x\nu_y}$ are given by
\begin{eqnarray}
A^{\nu_x\nu_y}_{N_x,y,5;1,y,1} &=& (-1)^{\nu_x} \quad \mbox{ for all }y,\\
A^{\nu_x\nu_y}_{x,N_y,6;x,1,2} &=& (-1)^{\nu_y} \quad \mbox{ for all }x,
\label{Abc}
\end{eqnarray}
with $A^{\nu_x\nu_y}_{ij}=-A^{\nu_x\nu_y}_{ji}$.
From the arrows in Fig. \ref{fig:star-fisher} one sees that the value $0$ ($1$) for $\nu_x$,
$\nu_y$ implies periodic (anti-periodic) boundary conditions on the matrix elements.
On the torus one then has\cite{k-61} (see also Ref. \onlinecite{wu-wang-08})
\be
\mathcal{Z}= \frac{1}{2}\sum_{\nu_x,\nu_y=0,1}r_{\nu_x\nu_y}\sqrt{\det A^{\nu_x\nu_y}},
\label{Ztdef}
\ee
where $r_{\nu_x\nu_y}$ are signs whose values will be determined below.

In order to calculate these determinants, as well as other
quantities to be considered later, we will use Samuel's reformulation
of the Pfaffian method in terms of a path integral description of noninteracting
Majorana fermions living on the lattice sites.\cite{samuel-1}
A fermion on site $i$ is associated with a Grassmann variable $\psi_i$. The
action $S$ of the fermions is defined in terms of a Kasteleyn
matrix $A$ of the dimer problem,
\be
S=\frac{1}{2}\sum_{ij}\psi_i A_{ij}\psi_j.
\label{S}
\ee
The partition function $Z$ of the fermions then equals the Pfaffian
of $A$,
\be
Z = \int {\cal D}\psi\, \exp(-S) = \mbox{Pf }A.
\label{Z-maj}
\ee
Here $\int {\cal D}\psi \equiv \int d\psi_1 \cdots d\psi_{N}$.
In the following the matrix $A$ is taken to be one of the four Kasteleyn matrices $A^{\nu_x\nu_y}$
for the dimer problem on the torus. To avoid cluttering the notation we will however not
write the superscripts explicitly. We first introduce new Grassmann variables $\tilde{\psi}_{\bm{k},\alpha}$
related to the $\psi_{\bm{r},\alpha}$ by a Fourier transformation,
\be
\psi_{\bm{r},\alpha}=\frac{1}{\sqrt{N_x N_y}}\sum_{\bm{k}}e^{i\bm{k}\cdot \bm{r}}\tilde{\psi}_{\bm{k},\alpha}.
\label{ft}
\ee
The form of the wavevectors $\bm{k}$ depends on the boundary conditions $\nu_x$ and $\nu_y$, i.e.
$k_x=2\pi(n_x + \nu_x/2)/N_x$ and similarly for $k_y$. The integers $n_x$ and $n_y$ are chosen
so that the sum over $\bm{k}$ runs over the first Brillouin zone. The partition function can now
be written $Z=\int {\cal D}\tilde{\psi}\,\exp(-S)$ with
\be
S=\frac{1}{2}\sum_{\bm{k}}\sum_{\alpha\beta}\tilde{\psi}_{-\bm{k},\alpha}\tilde{A}(\bm{k})_{\alpha\beta}
\tilde{\psi}_{\bm{k},\beta},
\label{Sk}
\ee
where $\tilde{A}(\bm{k})$ is the $6\times 6$ matrix
\be
\tilde{A}(\bm{k}) = \left(\begin{array}{cccccc}
0 & 1 & 1 & 0 & -e^{-ik_x} & 0 \\
-1 & 0 & 1 & 0 & 0 & -e^{-ik_y} \\
-1 & -1 & 0 & 1 & 0 & 0 \\
0 & 0 & -1 & 0 & 1 & 1 \\
e^{ik_x} & 0 & 0 & -1 & 0 & 1 \\
0 & e^{ik_y} & 0 & -1 & -1 & 0 \end{array}\right).
\ee
From (\ref{Sk}) it then follows that $\det A=Z^2$ is given by
\be
\det A = \prod_{\bm{k}}\det \tilde{A}(\bm{k}).
\ee
Because $\det \tilde{A}(\bm{k})=4$ for all $\bm{k}$, $\det A = 4^{N_x N_y} = 4^{N/6}$
independently of the boundary conditions $\nu_x$ and $\nu_y$.
Thus (\ref{Ztdef}) becomes $\mathcal{Z}=(r/2)\cdot 2^{N/6}$ where $r\equiv r_{00}+r_{01}+r_{10}+r_{11}$.
We expect $r$ to contribute to an $N$-independent term in the entropy (cf. Eqs. (\ref{ent-1})-(\ref{ent-2})).
Thus $r$ can be determined by manually counting the number of dimer coverings for a torus with only one unit
cell\cite{wu-wang-08} (i.e. $N=6$). We find $\mathcal{Z}=4$ for this case, which implies $r=4$, i.e. all terms in
(\ref{Ztdef}) come with a positive sign. Thus the number of dimer coverings on a star lattice on a torus with
$N$ sites is given by\cite{mis-sin-men}
\be
\mathcal{Z}=2^{N/6+1},
\label{Zt-result}
\ee
in agreement with the result (\ref{Zs}) derived from the arrow representation in Sec. \ref{ndc-arrows}.

\section{The Green function}
\label{green}

Within the framework of the fermionic formulation of the Pfaffian method,\cite{samuel-1}
quantities of interest for the dimer problem can be expressed in terms of
fermionic correlation functions.\cite{samuel-1,fms-02,iif-02,wang-wu-08}
The average number of dimers on a bond is given directly as the two-point Green
function of the Majorana fermions. Furthermore, dimer, vison, and monomer correlation
functions can be expressed in terms of multi-point Green functions, which in turn,
by Wick's theorem, are given as sums of products of two-point Green functions.
In this section we calculate the two-point Green function of the star-lattice
CDM for the case of equal dimer weights.

The expectation value of an arbitrary operator $O$ in the fermionic system defined by
(\ref{S}) and (\ref{Z-maj}) is $\langle O\rangle = Z^{-1}\int {\cal D}\psi\;O\,\exp(-S)$.
The two-point Green function (henceforth just called the Green function) is
then
\be
G_{ij}\equiv \langle \psi_i \psi_j\rangle = (A^{-1})_{ij}.
\label{gf-def}
\ee
Thus, viewed as a matrix, the Green function is the inverse of the Kasteleyn matrix:
$G=A^{-1}$. Using Eq. (\ref{ft}), one can write $G_{\bm{r},\alpha;\bm{r}',\alpha'}=\langle \psi_{\bm{r},\alpha}
\psi_{\bm{r}',\alpha'}\rangle=(N_x N_y)^{-1}\sum_{\bm{k},\bm{k}'}e^{i(\bm{k}\cdot\bm{r}+\bm{k}'\cdot\bm{r}')}
\langle \tilde{\psi}_{\bm{k},\alpha}\tilde{\psi}_{\bm{k}',\alpha'}\rangle$. From Eq. (\ref{Sk}) one finds
$\langle \tilde{\psi}_{\bm{k},\alpha}\tilde{\psi}_{\bm{k}',\alpha'}\rangle=\delta_{\bm{k}',-\bm{k}}
\tilde{A}^{-1}(\bm{k})_{\alpha\alpha'}$, where $\tilde{A}^{-1}(\bm{k})$ is the inverse matrix of $\tilde{A}(\bm{k})$.
Thus
\be
G_{\bm{r},\alpha;\bm{r}',\alpha'}=\int_{-\pi}^{\pi}\frac{dk_x}{2\pi}\int_{-\pi}^{\pi}\frac{dk_y}{2\pi}
e^{i\bm{k}\cdot(\bm{r}-\bm{r}')}\tilde{A}^{-1}(\bm{k})_{\alpha\alpha'}
\label{G-exp}
\ee
where we have also taken the thermodynamic limit $N_x$, $N_y\to \infty$.
In this limit the Green function becomes independent of the boundary conditions $\nu_x$, $\nu_y$
because the wavevectors are no longer discrete. The inverse matrix of $\tilde{A}(\bm{k})$ takes the form
\be
\tilde{A}^{-1}(\bm{k})=\frac{1}{4}\left(\begin{array}{cc} P & -R^{\dagger} \\ R & Q \end{array}\right)
\label{At-inv}
\ee
where $P$, $Q$, and $R$ are $3\times 3$ matrices given by
\begin{widetext}
\begin{eqnarray}
P &=& \left(\begin{array}{ccc}
e^{ik_y}-e^{-ik_y} & -1-e^{-ik_x}-e^{ik_y}+e^{-ik_x+ik_y} & -1+e^{-ik_x}-e^{-ik_y}-e^{-ik_x+ik_y} \\
1+e^{ik_x}+e^{-ik_y}-e^{ik_x-ik_y} & -e^{ik_x}+e^{-ik_x} & -1-e^{-ik_x}+e^{-ik_y}-e^{ik_x-ik_y} \\
1-e^{ik_x}+e^{ik_y}+e^{ik_x-ik_y} & 1+e^{ik_x}-e^{ik_y}+e^{-ik_x+ik_y} & e^{ik_x-ik_y}-e^{-ik_x+ik_y} \end{array}
\right),\label{P}\\
Q &=& \left(\begin{array}{ccc}
-e^{ik_x-ik_y}+e^{-ik_x+ik_y} & -1+e^{-ik_x}-e^{-ik_y}-e^{-ik_x+ik_y} & -1-e^{-ik_x}+e^{-ik_y}-e^{ik_x-ik_y} \\
1-e^{ik_x}+e^{ik_y}+e^{ik_x-ik_y} & -e^{ik_y}+e^{-ik_y} & -1-e^{ik_x}-e^{-ik_y}+e^{ik_x-ik_y} \\
1+e^{ik_x}-e^{ik_y}+e^{-ik_x+ik_y} & 1+e^{-ik_x}+e^{ik_y}-e^{-ik_x+ik_y} & e^{ik_x}-e^{-ik_x} \end{array}\right),
\label{Q}\\
R &=& \left(\begin{array}{ccc}
1+e^{ik_x}+e^{ik_y}-e^{ik_x-ik_y} & -1-e^{ik_x}-e^{ik_y}+e^{-ik_x+ik_y} & 2-e^{ik_x-ik_y}-e^{-ik_x+ik_y} \\
-2e^{ik_x}+e^{ik_x+ik_y}+e^{ik_x-ik_y} & 1+e^{ik_x}+e^{ik_y}-e^{ik_x+ik_y} & -1-e^{ik_x}-e^{ik_y}+e^{ik_x-ik_y} \\
1+e^{ik_x}+e^{ik_y}-e^{ik_x+ik_y} & -2e^{ik_y}+e^{ik_x+ik_y}+e^{-ik_x+ik_y} & 1+e^{ik_x}+e^{ik_y}-e^{-ik_x+ik_y}
\end{array}\right)\label{R}.
\end{eqnarray}
\end{widetext}
Since the dependence on $\bm{r},\bm{r}'$ in Eq. (\ref{G-exp}) is only through $\bm{r}-\bm{r}'\equiv \bm{R}$,
we define $G_{\bm{r},\alpha;\bm{r}',\alpha'}\equiv G(\bm{R})_{\alpha\alpha'}=G(R_x,R_y)_{\alpha\alpha'}$.
We see from Eqs. (\ref{At-inv})-(\ref{R}) that each term in the matrix elements of $\tilde{A}^{-1}(\bm{k})$ contains
at most one power of $e^{\pm i k_x}$ and at most one power of $e^{\pm i k_y}$. From Eq. (\ref{G-exp}) it then follows
that
\be
G(R_x,R_y)_{\alpha\alpha'}=0 \quad \mbox {if }|R_x|\geq 2 \mbox{ or if }|R_y|\geq 2,
\label{G-zero}
\ee
i.e. the Green function vanishes identically beyond a very short distance. The same property has recently also been
found for the Green function on the kagome lattice.\cite{wang-wu-07,wang-wu-08}

For each $\bm{R}$, $G(\bm{R})=G(R_x,R_y)$ defines a $6\times 6$ matrix. Of these, due to Eq. (\ref{G-zero}),
only $G(0,0)$, $G(1,0)$, $G(-1,0)$, $G(0,1)$, $G(0,-1)$, $G(-1,1)$, $G(1,-1)$, $G(1,1)$ and $G(-1,-1)$ can be nonzero.
However, not all of these matrices are independent because $G=A^{-1}$ is antisymmetric and thus
$G(-R_x,-R_y)=-G(R_x,R_y)^T.$

The results for the Green function obtained in this section will be used in the fermionic calculations of the
quantities considered in Secs. \ref{d-occ}-\ref{mm-corr}.

\section{Dimer occupation probabilities}
\label{d-occ}

Let $i$ and $j$ be two sites connected by a bond. Since for any dimer covering of the lattice, the bond $ij$ is
either occupied by a dimer or not, the number of dimers $n_{ij}$ on the bond can
only take the values 0 or 1. We define $p(ij)=\langle n_{ij}\rangle$ as the average of $n_{ij}$ over all dimer
coverings. Thus $p(ij)$ is the probability that bond $ij$ is occupied by a dimer. For the star lattice with
the same dimer weight on all bonds, symmetry dictates that $p(ij)$ may take at most two different values
as the bond $ij$ is varied: $p(ij)=p_t$ if $ij$ is a triangle bond and $p(ij)=p_l$ if $ij$ is a linking bond.
Explicit calculations give
\be
p_t=\frac{1}{4}, \quad p_l=\frac{1}{2}
\label{dimer-occ-res}
\ee
Before turning to the derivation of this result, let us show that $p_t$ and $p_l$ are
related to each other. This can be seen by considering the average number of dimers in
the whole system, given by $\langle \sum_{(ij)}n_{ij}\rangle = N p_t + (N/2)p_l$. Here the sum goes over
all bonds and the prefactors of $p_t$ and $p_l$ are the number of triangle bonds and linking bonds, respectively,
in a system with $N$ sites. The average is trivially equal to $N/2$ since each dimer covering satisfies the
close-packed hard-core constraint. Thus
\be
p_t + \frac{1}{2}p_l = \frac{1}{2}.
\label{n-cond}
\ee

\subsection{Fermionic approach}

The probability $p(ij)$ for having a dimer on bond $ij$ is given by\cite{unit-weight-form}
\be
p(ij) = \langle n_{ij}\rangle= -\langle \psi_i \psi_j\rangle = -G_{ij}.
\label{pij-Gij}
\ee
In this expression the site labels $i$ and $j$ have been chosen such that $A_{ij}>0$, i.e.
the Kasteleyn arrow goes from $i$ to $j$. Eq. (\ref{pij-Gij}) can be derived\cite{wang-wu-08}
from the form of the generating function for dimer coverings when the dimer weights are allowed
to be arbitrary (the dimer generating function reduces to our $\mathcal{Z}$ when all dimer weights
are equal). The result (\ref{dimer-occ-res}) now easily follows by using Eqs. (\ref{G-exp})-(\ref{R}) to evaluate
Eq. (\ref{pij-Gij}).

\begin{figure}[!htb]
\begin{center}
\centerline{\includegraphics[scale=0.6]{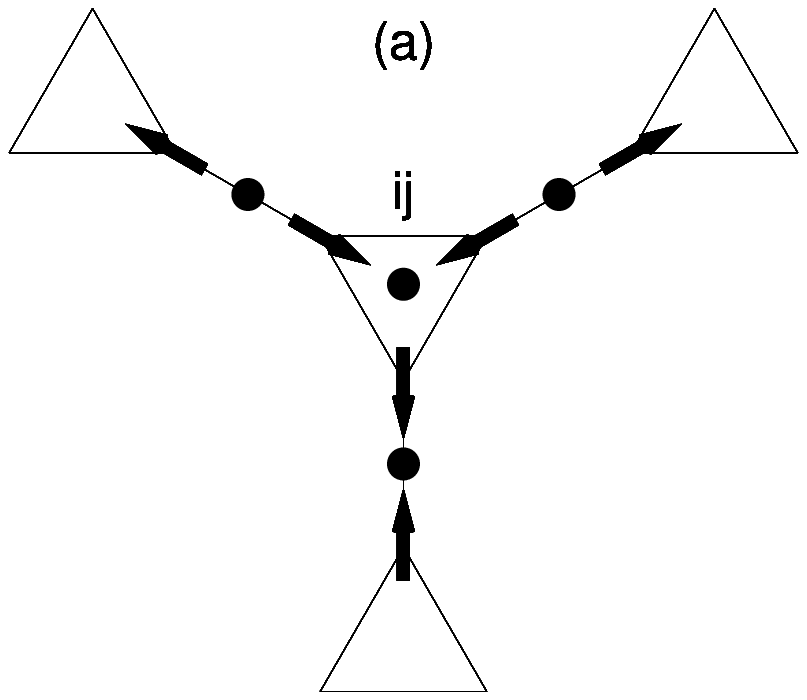}\hspace{1cm}
\includegraphics[scale=0.6]{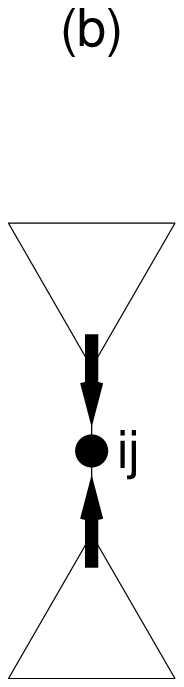}}
\caption{The presence of a dimer on bond $ij$ leads to a set of ``spent'' arrows and
constraints as shown (the filled circles symbolize the constraints). This collection
of arrows and constraints is referred to as the dimer's imprint. (a) Imprint of a
dimer on a triangle bond $ij$. (b) Imprint of a dimer on a linking bond $ij$. The
probability $p(ij)$ of having a dimer on the bond $ij$ is found by inserting the
number of spent arrows and constraints $\Delta \mathcal{N}_a$ and
$\Delta \mathcal{N}_c$ into Eq. (\ref{rat-2}).}
\label{fig:ratio-arrows}
\end{center}
\end{figure}

\subsection{Arrow approach}
\label{ratio-arrow}

The probability $p(ij)=\langle n_{ij}\rangle$ of having a dimer on bond $ij$ is given by the ratio
\be
p(ij) = \frac{\mathcal{Z}(\{n_{ij}=1\})}{\mathcal{Z}}
\label{nij-ratio}
\ee
where $\mathcal{Z}(\{n_{ij}=1\})$ is the number of dimer coverings in which bond $ij$ is
occupied by a dimer and $\mathcal{Z}$ is the total number of dimer coverings. In the
sections that follow we will discuss other quantities that can be written in this way as
well, i.e. as the ratio of a numerator, call it $\mathcal{Z}'$ in general, and the
denominator which is $\mathcal{Z}$ in all cases. In Sec. \ref{ndc-arrows} the latter was
expressed as $\mathcal{Z}=2^{{\cal N}_a-{\cal N}_c}$ where ${\cal N}_a$ is the number
of flippable arrows (which are Ising (i.e. $Z_2$) degrees of freedom) and ${\cal N}_c$ is
the number of independent constraints on these arrows. For all quantities of interest, the
numerator $\mathcal{Z}'$ can be expressed in terms of arrows in a completely analogous way,
the only difference being that the calculation of $\mathcal{Z}'$ corresponds to imposing
additional requirements on some of these arrows. For example, some arrows may be required
to point in definite directions. If these additional requirements are incompatible with
the arrow constraints it follows that the numerator vanishes. If, on the other hand,
they are compatible with the constraints, they will cause the number of remaining
constraints and flippable arrows to change by $\Delta {\cal N}_c$ and $\Delta {\cal N}_a$,
respectively, compared to their values in the calculation of $\mathcal{Z}$. This gives the result
\be
\frac{\mathcal{Z}'}{\mathcal{Z}}=2^{\Delta {\cal N}_a -\Delta {\cal N}_c}.
\label{rat-2}
\ee
The calculation of such a ratio therefore amounts to identifying the values
of $\Delta {\cal N}_a$ and $\Delta {\cal N}_c$ for the quantity of interest.

Having explained the calculation in general, let us now apply this arrow approach
to Eq. (\ref{nij-ratio}), for which the dimer coverings that contribute to the
numerator are characterized by having a dimer on bond $ij$. Let us first consider
the case of $ij$ being a triangle bond. The presence of a dimer on
such a bond implies that the arrows on site $i$ and site $j$
must both point into the associated triangle. The arrow constraint for the triangle
then implies that the arrow on the third site of the triangle must point out of the
triangle. Next, as each of the three fixed arrows on the triangle also belongs to a
linking bond, the constraint on each of the three linking bonds fixes the direction
of the arrow on the other end of the linking bond. Thus the directions of six arrows
become fixed by having a dimer on the triangle bond $ij$. These arrows are therefore
no longer flippable, Ising degrees of freedom, so $\Delta \mathcal{N}_a=-6$.
Furthermore, four constraints are also spent (the triangle constraint and three linking
bond constraints) so $\Delta \mathcal{N}_c=-4$. This gives $p(ij) = 2^{-6-(-4)}=1/4=p_t$.
This case is illustrated in Fig. \ref{fig:ratio-arrows}(a).

Next consider the case of $ij$ being a linking bond. Having a dimer on such a
bond implies that the two arrows must point into (towards the center of) the
linking bond, giving $\Delta {\cal N}_a=-2$. Furthermore, $\Delta {\cal N}_c=-1$
since the constraint on that linking bond is then spent. Thus $p(ij) =
2^{-2-(-1)}=1/2=p_l$. This case is illustrated in Fig. \ref{fig:ratio-arrows}(b).

The set of spent arrows and constraints that results from having a dimer on
a given bond will be referred to as the dimer's \textit{imprint}. Fig. \ref{fig:ratio-arrows}(a)
and (b) shows the dimer imprint for a triangle-bond dimer and a linking-bond dimer, respectively.
Thus from counting the number of arrows and constraints in the imprint of a dimer on the bond
$ij$, one can use Eq. (\ref{rat-2}) to calculate the probability $p(ij)$ of having a dimer
on that bond.

\section{Dimer correlations}
\label{d-corr}

In this section we consider correlations between dimers in the star-lattice
CDM with equal dimer weights. The main focus is on correlations between two
dimers, but at the end we also briefly discuss correlations between three dimers.

\begin{figure}[!htb]
\begin{center}
\centerline{\includegraphics[scale=0.6]{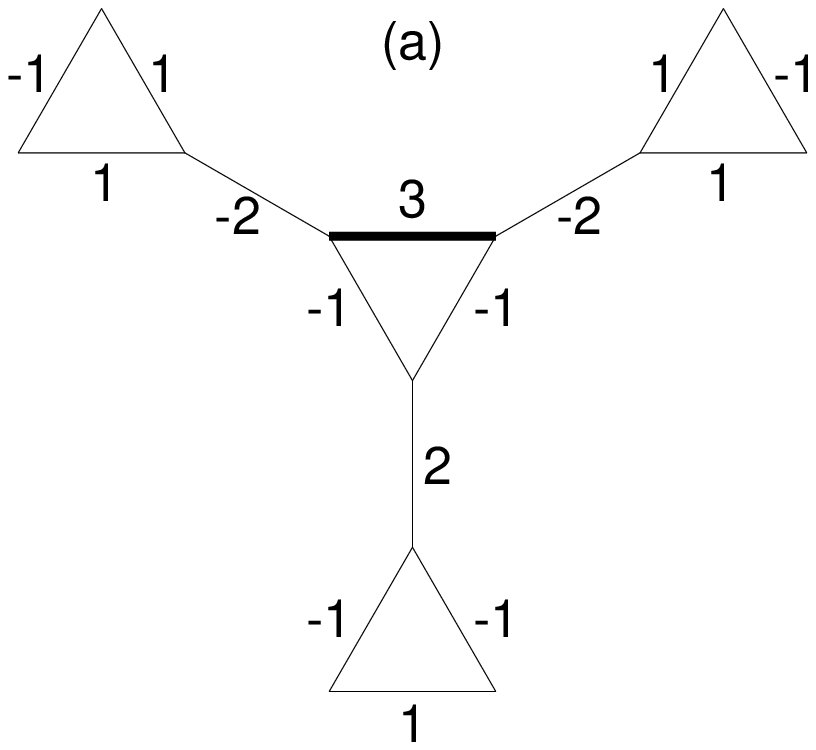}\hspace{1cm}
\includegraphics[scale=0.6]{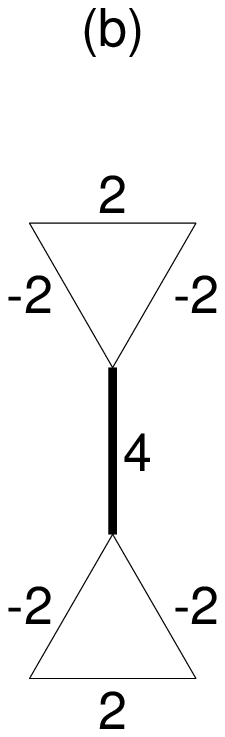}}
\caption{Dimer-dimer correlations on the star lattice.
The number next to a bond $kl$ is the dimer-dimer correlation function $c(ij,kl)$
(in units of $1/16$) where the reference bond $ij$ is shown in bold. (a)
The reference bond is a triangle bond. (b) The reference bond is a linking bond.
In both (a) and (b) only bonds $kl$ with $c(ij,kl)\neq 0$ are shown. Note that
the correlations between dimers on different bond types can be read off from
both (a) and (b).}
\label{fig:dd-corr}
\end{center}
\end{figure}

The probability that bonds $ij$ and $kl$ both have a dimer is given by
$p(ij,kl)=\langle n_{ij}n_{kl}\rangle$. The correlations between
dimers on bonds $ij$ and $kl$ are quantified by the dimer-dimer correlation
function\cite{fs-63}
\begin{eqnarray}
c(ij,kl) &=& p(ij,kl)-p(ij)p(kl)\nonumber \\ &=&
\langle n_{ij}n_{kl}\rangle - \langle n_{ij}\rangle \langle n_{kl}\rangle.
\label{c-def}
\end{eqnarray}
This quantity can have either sign and vanishes for uncorrelated dimers. Our
results for $c(ij,kl)$ are presented in Fig. \ref{fig:dd-corr} (all numbers are
in units of $1/16$). We find that a dimer is uncorrelated with all dimers located
further away than on one of the closest triangles. In particular, dimers on
linking bonds are uncorrelated with each other.

\subsection{Fermionic approach}

In the fermionic approach $\langle n_{ij}n_{kl}\rangle$ is given by a four-point correlation
function,\cite{unit-weight-form}
\be
\langle n_{ij}n_{kl}\rangle = \langle \psi_i \psi_j \psi_k \psi_l\rangle.
\label{nijkl-fermi}
\ee
This expression for $\langle n_{ij}n_{kl}\rangle$ can be derived from the generating function
of dimer coverings in the same way as Eq. (\ref{pij-Gij}).\cite{wang-wu-08} (In this derivation
the site labels were chosen so that the two Kasteleyn arrows
go from $i$ to $j$ and from $k$ to $l$, respectively; cf. the remark after Eq. (\ref{pij-Gij})).
Since the fermionic action is quadratic, the four-point function $\langle\psi_i\psi_j\psi_k\psi_l\rangle$ can be
evaluated using Wick's theorem. Inserting the Wick expansion into Eq. (\ref{c-def}) and using
Eq. (\ref{pij-Gij}), the term $G_{ij}G_{kl}$ cancels, giving
\be
c(ij,kl)=G_{il}G_{jk}-G_{ik}G_{jl}.
\label{d-corr-Gf}
\ee
The property (\ref{G-zero}) of the Green function then implies that also the dimer-dimer correlation
function will vanish beyond a very short distance. (Note that Eq. (\ref{nijkl-fermi}) is not valid
for the case $ij=kl$ when one instead has $\langle n_{ij}^2\rangle = -G_{ij}$, since $n_{ij}^2=n_{ij}$. Thus
$c(ij,ij)=-G_{ij}-G_{ij}^2$.) The results from the Green function calculation of the dimer-dimer
correlation function are shown in Fig. \ref{fig:dd-corr}.

\subsection{Arrow approach}

The probability $p(ij,kl)=\langle n_{ij}n_{kl}\rangle$ is given by
\be
p(ij,kl)=\frac{\mathcal{Z}(\{n_{ij}=1,n_{kl}=1\})}{\mathcal{Z}}
\label{pijkl-ratio}
\ee
where $\mathcal{Z}(\{n_{ij}=1,n_{kl}=1\}$ is the number of dimer coverings with
dimers on both $ij$ and $kl$. Eq. (\ref{pijkl-ratio}) is a ratio of the type (\ref{rat-2}) that
can be evaluated with the arrow representation as explained in Sec. \ref{ratio-arrow}.
One starts by laying down the imprints of spent arrows and constraints for each of the
two dimers (these imprints are shown for the two types of dimers in Fig. \ref{fig:ratio-arrows}).
If the dimers are sufficiently far apart that their respective imprints have no arrows
in common, the two dimers contribute additively to $\Delta \mathcal{N}_a-\Delta \mathcal{N}_c$.
This gives $p(ij,kl)=p(ij)p(kl)$, i.e. the dimers are uncorrelated.\cite{additional-ac}
In contrast, dimers are correlated if their imprints overlap. In this situation $p(ij,kl)=0$ if the overlapping
parts do not match, which happens if the two dimers cannot both be present in a dimer covering.
Alternatively, if the overlapping parts do match, the value of $\Delta \mathcal{N}_a-\Delta \mathcal{N}_c$
is larger (i.e., less negative) than the sum of the contributions from each dimer considered separately, thus giving
$p(ij,kl)> p(ij)p(kl)$. This arrow calculation reproduces the results for $c(ij,kl)$ shown
in Fig. \ref{fig:dd-corr}. Note that a dimer on a triangle bond (Fig. \ref{fig:dd-corr}(a))
is correlated with dimers further away than a dimer
on a linking bond is (Fig. \ref{fig:dd-corr}(b)). This difference is a consequence of the different sizes
of their respective imprints shown in Fig. \ref{fig:ratio-arrows}. In particular, the imprint of linking
bond dimers is so small that such dimers are only correlated with dimers on the two triangles
touching the linking bond. Dimers on triangle bonds, having a somewhat bigger imprint, are correlated
only if they are on the same or neighboring triangles.

The notion of an imprint can be generalized to more than one dimer. For example, the imprint of two
dimers at $ij$ and $kl$ is given by the collection of spent arrows and constraints resulting from those
two dimers. If the arrows are mutually compatible, the probability $p(ij,kl)$ can be found by inserting
the number of spent arrows and constraints into Eq. (\ref{rat-2}). Another example will be given
in the next subsection.

\subsection{Correlations between three dimers}
\label{ddd-corr}

\begin{figure}[!htb]
\begin{center}
\centerline{\includegraphics{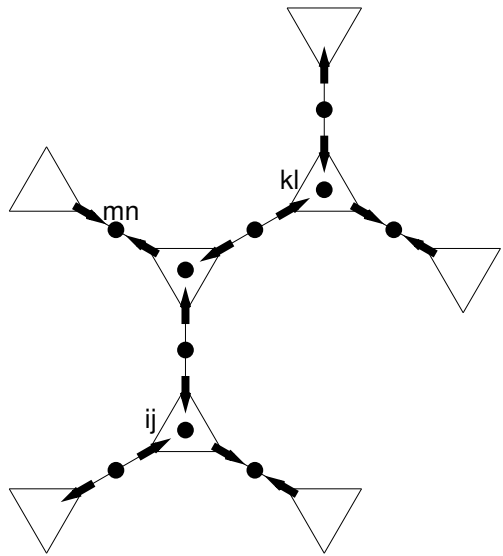}}
\caption{Imprint of spent arrows and constraints for $p(ij,kl,mn)$ where the
three bonds $ij$, $kl$, and $mn$ are as shown. This imprint is the same as that for $p(ij,kl)$.
It has $\Delta \mathcal{N}_a=-14$ and $\Delta \mathcal{N}_c=-10$, giving $p(ij,kl,mn)=p(ij,kl)=1/16$.}
\label{fig:ddd-corr}
\end{center}
\end{figure}

Correlations between three (or more) dimers can be similarly analyzed. Let us briefly consider
the three-dimer case. This will involve the probability $p(ij,kl,mn)=\langle n_{ij}n_{kl}n_{mn}\rangle$
that the bonds $ij$, $kl$, and $mn$ are all occupied by a dimer. Note that even if all three
dimers are pairwise uncorrelated, i.e. $p(ij,kl)=p(ij)p(kl)$, $p(ij,mn)=p(ij)p(mn)$,
and $p(kl,mn)=p(kl)p(mn)$, it isn't necessarily true that $p(ij,kl,mn)$ equals
$p(ij)p(kl)p(mn)$. As an example, consider the situation with the bonds $ij$, $kl$, and $mn$ as
shown in Fig. \ref{fig:ddd-corr}. It follows from the previous analysis in this
section that dimers on these three bonds are indeed all pairwise uncorrelated.
However, it is easy to see that if there is a dimer on both $ij$ and $kl$, there must also
be a dimer on $mn$. Thus $p(ij,kl,mn)=p(ij)p(kl)=1/16$ in this case. This result
is also found from the fermionic approach, using that\cite{unit-weight-form}
\be
\langle n_{ij}n_{kl}n_{mn}\rangle = -\langle \psi_i \psi_j \psi_k \psi_l \psi_m \psi_n\rangle
\label{nijklmn-fermi}
\ee
which is then evaluated using Wick's theorem (Eq. (\ref{nijklmn-fermi}) can be
derived in the same way as Eqs. (\ref{pij-Gij}) and (\ref{nijkl-fermi})).
Alternatively, one can use the arrow representation. The imprint of spent arrows and
constraints for $p(ij,kl,mn)$ is shown in Fig. \ref{fig:ddd-corr}. The fact that
this is also the imprint for $p(ij,kl)$ shows that $p(ij,kl,mn)=p(ij,kl)$. From this figure
one sees that $\Delta \mathcal{N}_a=-14$ and $\Delta \mathcal{N}_c=-10$, and thus
$p(ij,kl,mn)=2^{-14-(-10)}=1/16$.

\begin{figure}[!htb]
\begin{center}
\centerline{\includegraphics[scale=0.65]{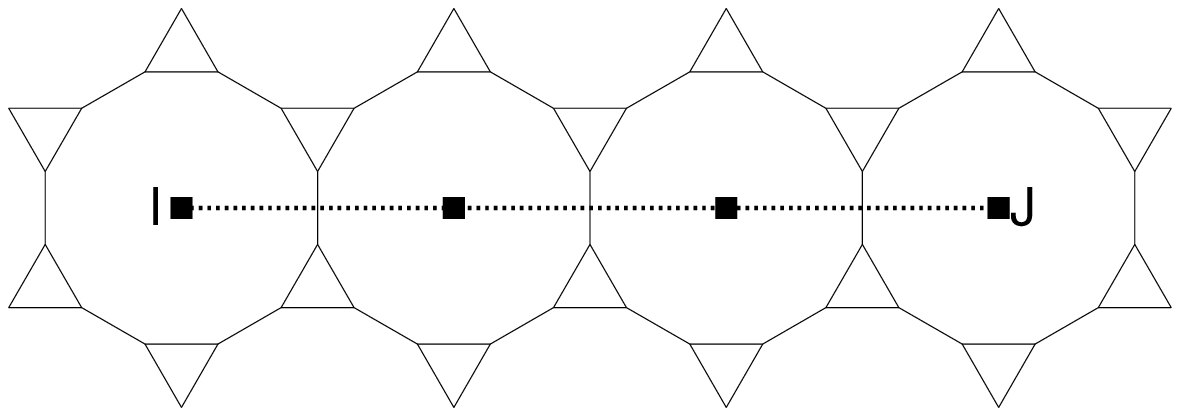}}
\centerline{\includegraphics[scale=0.65]{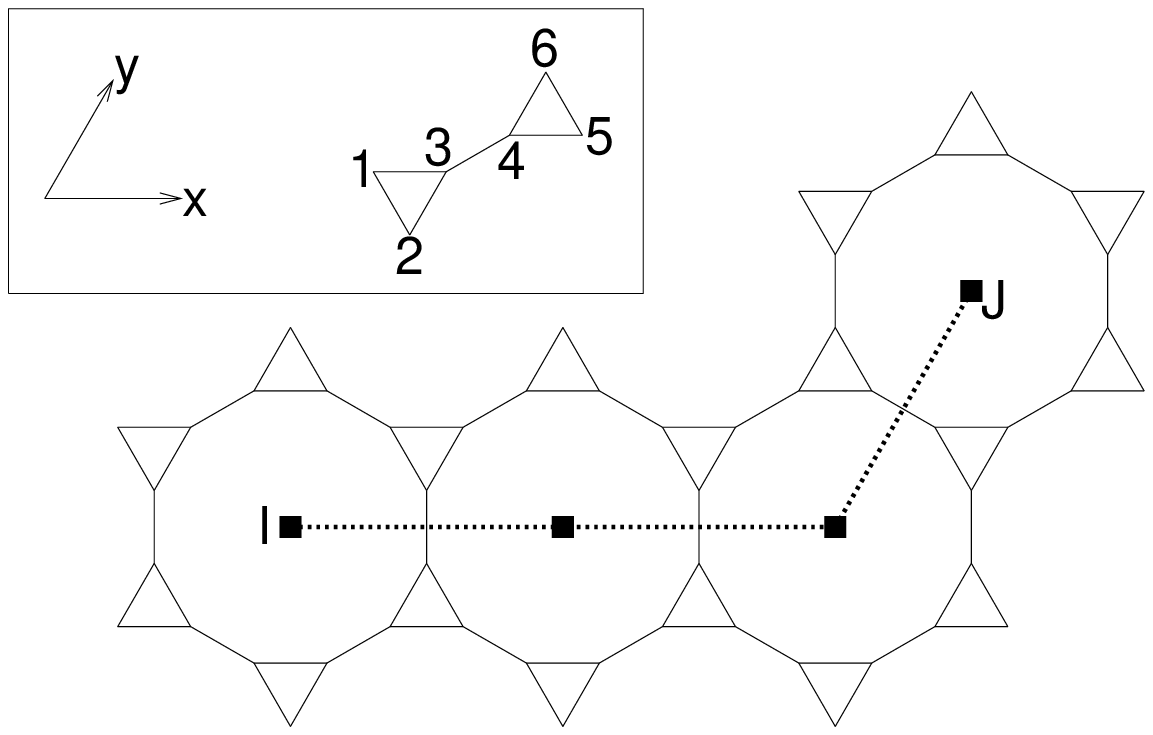}}
\caption{Two examples (top and bottom) of a pair of visons on the star lattice.
The two visons are located at the dodecagons whose centers are sites $I$ and $J$ on the dual lattice.
A path $\Gamma$ on the dual lattice (dotted line) connects the two vison sites.
The vison-vison correlation function is equal to $|f_{\rm{e}}-f_{\rm{o}}|$ where $f_{\rm{e}}$ ($f_{\rm{o}}$)
is the fraction of all dimer coverings that has an even (odd) number of dimers on the $n$ bonds
intersected by $\Gamma$. The top figure is an example of the case considered in Sec. \ref{vv-fermi}, with
$n(=n_x)=3$. The bottom figure is an example of the case considered in Sec. \ref{vv-arrow}, with
$n_x=2$, $n_y=1$ and $n=3$. The inset shows the axes and unit cell in Fig. \ref{fig:star-fisher}
translated to this figure.}
\label{fig:visons}
\end{center}
\end{figure}

\section{Vison correlations}
\label{vv-corr}

The excited-state manifold corresponding to the lowest excitation energy of the
QDM Hamiltonian (\ref{qdm-ham}) is spanned by states that have
eigenvalue $-1$ for $\hat{\sigma}^x(D)$ on two of the dodecagons and $+1$ for
all the others. The two localized $\sigma^x(D)=-1$ excitations are called (point)
visons. A vison can thus be labeled by the center of the hosting dodecagon.
Let us find an expression for the state $|\Psi_{I,J}\rangle$ (in a given topological
sector) with visons at the dodecagon sites $I$ and $J$ (we use capital letters
for sites on the dual lattice; the dodecagon centers form a subset of all dual
lattice sites). The reasoning is analogous to the one used for the kagome
lattice.\cite{mis-kag-1} Consider a path $\Gamma$ on the dual lattice that connects
sites $I$ and $J$, as shown in Fig. \ref{fig:visons}. For an arbitrary dimer
covering $|c\rangle$, let the operator $\hat{N}_{\Gamma}$ count
the number of dimers $N_{\Gamma}(c)$ on the bonds intersected by $\Gamma$. Then the
operator $(-1)^{\hat{N}_{\Gamma}}$ measures the associated parity, i.e. whether
$N_{\Gamma}(c)$ is even or odd. The dimer configuration around a dodecagon $D$
corresponds to one of the $32$ even-length loops around the dodecagon
having an alternation of bonds with/without a dimer (cf. the discussion around Eq. (\ref{sigma-x})).
As $\hat{\sigma}^x(D)$ shifts the dimers by one bond along this loop, it can be seen that
$\hat{\sigma}^x(D)$ conserves the parity if $\Gamma$ crosses the loop an even number of times,
and flips the parity if $\Gamma$ crosses the loop an odd number of times.
The latter case is only realized if $D$ is one of the endpoints $I$
or $J$ of $\Gamma$. Thus $\hat{\sigma}^x(D) (-1)^{\hat{N}_{\Gamma}}=
(-1)^{\hat{N}_{\Gamma}}\hat{\sigma}^x(D)$ if $D\neq I,J$ and $\hat{\sigma}^x(D) (-1)^{\hat{N}_{\Gamma}}=
-(-1)^{\hat{N}_{\Gamma}}\hat{\sigma}^x(D)$ if $D=I,J$. It follows that
the two-vison state we are looking for is given by $|\Psi_{I,J}\rangle\equiv (-1)^{\hat{N}_{\Gamma}}|\mbox{RK}\rangle$
(here $|\mbox{RK}\rangle$ is the Rokhsar-Kivelson ground state in the given topological sector)
since it satisfies $\hat{\sigma}^x(D)|\Psi_{I,J}\rangle = |\Psi_{I,J}\rangle$ for $D\neq I,J$ and
$\hat{\sigma}^x(D)|\Psi_{I,J}\rangle = -|\Psi_{I,J}\rangle$ for $D=I,J$.
The vison-vison correlation function is equal to (the absolute value of) the
ground-state expectation value of $(-1)^{\hat{N}_{\Gamma}}$,
\be
v(I,J)\equiv |\langle \mbox{RK}|(-1)^{\hat{N}_{\Gamma}}|\mbox{RK}\rangle|.
\label{vv-def}
\ee
Clearly $v(I,J)=0$ since it is the overlap of $|\mbox{RK}\rangle$ and
$|\Psi_{I,J}\rangle$ which are eigenstates of the Hamiltonian with different
eigenvalues. As $(-1)^{\hat{N}_{\Gamma}}$ is diagonal in the dimer basis,
$v(I,J)$ can alternatively be expressed as the correlation function
\be
v(I,J)=|\langle (-1)^{N_{\Gamma}}\rangle|
\label{vv-classical}
\ee
in the classical dimer problem.\cite{fms-02,iif-02,misguich-mila-08} It is
instructive to rederive the vanishing of $v(I,J)$ from Eq. (\ref{vv-classical}),
which is what we consider in the remainder of this section.

\subsection{Fermionic approach}
\label{vv-fermi}

By changing the signs of the matrix elements $A_{ij}=-A_{ji}$ for bonds
$ij$ intersected by $\Gamma$ (which corresponds to reversing the direction
of the Kasteleyn arrows on these bonds), a dimer covering $|c\rangle$ is counted with
the sign $(-1)^{N_{\Gamma}(c)}$ as required by Eq. (\ref{vv-classical}).
In the fermionic formulation of this expectation value, these sign changes
are implemented by multiplying $\exp(-S)$ by a compensating factor in the
form of the ``string'' $\prod_{(ij)\in \Gamma}\exp(2\psi_i A_{ij}\psi_j)
=\prod_{(ij)\in \Gamma}(1+2\psi_i A_{ij}\psi_j)$, where the product runs
over all bonds intersected by $\Gamma$. Thus the vison-vison correlation
function involves the expectation value of this string,
\be
v(I,J) = \left|\left\langle \prod_{(ij)\in \Gamma}(1+2\psi_i A_{ij}\psi_j)\right\rangle\right|.
\label{vv-fermion}
\ee
We may choose $\Gamma$ to only visit dodecagon sites, so that all bonds intersected
are of the linking bond type.

As an example of the evaluation of Eq. (\ref{vv-fermion})
we will consider the simple case shown in Fig. \ref{fig:visons}(top), where $I$ and $J$ have
the same $y$ coordinate and are separated by a distance $n$, so that $\Gamma$ can be
taken as a straight line of length $n$ along the $x$ axis. Thus $\Gamma$ intersects $n$
linking bonds. Let $I=(X,Y)$ and $J=(X+n,Y)$ where $(X,Y)=(x+1/2,y+1/2)$.
Then $v(I,J)=v(X,Y;X+n,Y)\equiv |\tilde{v}(n)|$ where $\tilde{v}(n)$ is the expectation
value of the string. A direct evaluation of Eq. (\ref{vv-fermion}) gives $\tilde{v}(1)=\tilde{v}(2)=0$.
For $n\geq 3$ we write
\begin{widetext}
\be
\tilde{v}(n) = \left \langle \prod_{\ell=1}^n (1+2\psi_{x+\ell,y,6}\psi_{x+\ell,y+1,2})\right\rangle
= \tilde{v}(n-1) + 2 \left\langle \left[\prod_{\ell=1}^{n-1}(1+2 \psi_{x+\ell,y,6}\psi_{x+\ell,y+1,2})\right]
\psi_{x+n,y,6}\psi_{x+n,y+1,2}\right\rangle.
\label{vv-first}
\ee
Using Wick's theorem, the last expectation value is the sum of
$\tilde{v}(n-1)\langle \psi_{x+n,y,6}\psi_{x+n,y+1,2}\rangle$
and
\be
\left\langle \left[\prod_{\ell=1}^{n-2}(1+2 \psi_{x+\ell,y,6}\psi_{x+\ell,y+1,2})\right]
(1+2\psi_{x+n-1,y,6}\psi_{x+n-1,y+1,2})\psi_{x+n,y,6}\psi_{x+n,y+1,2}\right\rangle_{\rm{ext}}
\label{exp-pr-1}
\ee
where the subscript "ext" (external) means that when evaluating this expression using Wick's theorem, the last
two Grassmann variables (GV's) $\psi_{x+n,y,6}$ and $\psi_{x+n,y+1,2}$ should not be paired with each other.
Because of Eq. (\ref{G-zero}) any pairings
between GV's whose $x$ coordinates differ by 2 or more will vanish. It follows that the contribution to (\ref{exp-pr-1})
from the first term in $1+2\psi_{x+n-1,y,6}\psi_{x+n-1,y+1,2}$ will vanish. Furthermore, the only non-vanishing
pairings of $\psi_{x+n,y,6}$ and $\psi_{x+n,y+1,2}$ are with the GV's $\psi_{x+n-1,y,6}$ and $\psi_{x+n-1,y+1,2}$
of the neighboring bond. Thus (\ref{exp-pr-1}) is equal to
$2\tilde{v}(n-2)\langle \psi_{x+n-1,y,6}\psi_{x+n-1,y+1,2}\psi_{x+n,y,6}\psi_{x+n,y+1,2}\rangle_{\rm{ext}}$, so that
\begin{eqnarray}
\lefteqn{\tilde{v}(n) = \tilde{v}(n-1)[1+2\langle\psi_{x+n,y,6}\psi_{x+n,y+1,2}\rangle]} \nonumber \\ &+& 4 \tilde{v}(n-2)
[\langle\psi_{x+n-1,y,6}\psi_{x+n,y+1,2}\rangle \langle \psi_{x+n-1,y+1,2}\psi_{x+n,y,6}\rangle
-\langle\psi_{x+n-1,y,6}\psi_{x+n,y,6}\rangle   \langle \psi_{x+n-1,y+1,2}\psi_{x+n,y+1,2}\rangle].
\label{vn-final}
\end{eqnarray}
\end{widetext}
The coefficients of $\tilde{v}(n-1)$ and $\tilde{v}(n-2)$ in this expression vanish, so
$\tilde{v}(n)=0$ also for $n\geq 3$. (In fact, we may use Eq. (\ref{vn-final}) to conclude that
$\tilde{v}(n)=0$ even without evaluating these coefficients, as it follows from an induction proof,
using $\tilde{v}(1)=\tilde{v}(2)=0$). Thus $v(X,Y;X+n,Y)=|\tilde{v}(n)|=0$ for all $n$.

\subsection{Arrow approach}
\label{vv-arrow}

In this subsection we will use arguments based on the arrow representation to consider the vison-vison correlation
function $v(X,Y;X+n_x,Y+n_y)$ with both $n_x$ and $n_y$ arbitrary. In this case $\Gamma$ can be taken to be made
up of two straight line segments of length $|n_x|$ in the $x$ direction and length $|n_y|$ in the $y$ direction.
An example with $n_x$ and $n_y$ positive is shown in Fig. \ref{fig:visons}(bottom). Again, we note that all
the bonds intersected by $\Gamma$ are of the linking bond type. The fraction of dimer coverings that has a specific
distribution of dimers on the $n$ intersected bonds ($n=|n_x|+|n_y|$) is equal to $2^{-n}$. This is because
the presence or absence of a dimer on a linking bond has a probability $p_t=1-p_t=1/2$, and dimers can
be distributed independently on the linking bonds intersected by $\Gamma$. This independence can be seen by laying
down arrows corresponding to a particular dimer distribution on these linking bonds; the bonds will contribute
additively to $\Delta \mathcal{N}_a-\Delta \mathcal{N}_c$. (Note that for a linking bond, not only the presence, but
also the absence of a dimer, is associated with a definite configuration of the arrows on the two sites
connected by that linking bond.)
Distributing dimers on the intersected bonds in all possible ways should generate all dimer coverings. To
check that our arguments produce this result, we note that the number of ways to lay down $k$ dimers on $n$
bonds is ${n \choose k}$ and thus the total
fraction of configurations generated this way is $2^{-n}\sum_{k=0}^{n}{n \choose k} = 1$ as expected. Furthermore,
the fraction of configurations $f_{\rm{e}}$ and $f_{\rm{o}}$ with an even and odd number of intersected dimers,
respectively, is given by
\begin{eqnarray}
f_{\rm{e}} &=& 2^{-n}\sum_{k \rm{\;even} \atop 0\leq k \leq n}{n \choose k} = \frac{1}{2}, \label{fe}\\
f_{\rm{o}} &=& 2^{-n}\sum_{k \rm{\;odd} \atop 0\leq k \leq n} {n \choose k} = \frac{1}{2}.\label{fo}
\end{eqnarray}
Since the vison-vison correlation function (\ref{vv-classical}) can be written $|f_{\rm{e}}-f_{\rm{o}}|$,
Eqs. (\ref{fe})-(\ref{fo}) imply that it vanishes.

\section{Monomer correlations}
\label{mm-corr}

All properties considered so far have been for a system in which
every site is touched by exactly one dimer. A different but closely
related problem involves a lattice in which some sites are occupied
by monomers, meaning that those sites cannot be touched by a dimer.
In this section we consider a system with two monomers.
We will be interested in the monomer-monomer correlation function,
defined as
\be
m(i,j)=\frac{\mathcal{Z}_{mm}(i,j)}{\mathcal{Z}}
\label{mm-corr-def}
\ee
where, for the CDM with equal dimer weights, $\mathcal{Z}_{mm}(i,j)$ is the number of possible dimer coverings
when the system has monomers at sites $i$ and $j$,\cite{dimer-monomer} and
$\mathcal{Z}$ is the same as before, i.e. the number of dimer coverings for
the system in the absence of monomers. More generally, if not all dimer weights
are equal, the two quantities in Eq. (\ref{mm-corr-def}) are the generating
functions for the system with and without the monomers.\cite{fs-63}

The monomers in a CDM can be characterized as confined or deconfined. A deconfined
phase is characterized by a nonzero value of $m(i,j)$ in the limit of infinite monomer
separation $|i-j|\to\infty$, while in a confined phase $m(i,j)$ goes to zero in
this limit. Using Fisher's mapping,\cite{f-66} Moessner and Sondhi\cite{ms-03} were
able to show that the star-lattice CDM has a phase transition as a function of the
dimer weights in which monomers are confined in one phase and deconfined in the other.
They did this by relating certain sums of four monomer-monomer correlation functions
to a spin-spin correlation function in the square-lattice Ising model. This approach
does however not allow one to deduce $m(i,j)$ itself.

In this section we calculate the monomer-monomer correlation function for
a star-lattice CDM with equal dimer weights. We find that $m(i,j)=1/4$ for
all possible choices of the monomer sites $i$ and $j$ (with $i\neq j$) except when
$i$ and $j$ are connected by a linking bond, in which case $m(i,j)=1/2$. Thus
the monomers are deconfined in this case. As we discuss in Sec. \ref{mon-Ising},
these results are consistent with the analysis in Ref. \onlinecite{ms-03}.

\subsection{Fermionic approach}

\begin{figure*}[htb]
\begin{center}
\centerline{\includegraphics[scale=1.3]{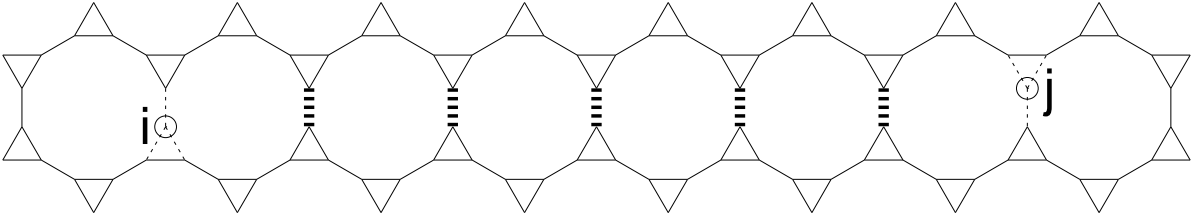}}
\caption{A particular case of the two-monomer system for which Eq. (\ref{mij-fermi}) for the monomer-monomer
correlation function $m(i,j)$ is evaluated in the text. The two monomers are located at sites $i$ and $j$ of
the original lattice (shown as open circles). In this figure, $i=(x,y,6)$ and $j=(x+n,y+1,2)$ with $n=6$
(cf. the site labeling conventions in Fig. \ref{fig:star-fisher} and the inset in Fig. \ref{fig:visons}).
The removed bonds connected to these sites are shown as dashed lines. The bonds whose Kasteleyn arrows are
reversed in this fermionic calculation are shown as thick dotted lines.}
\label{fig:mm-fermion}
\end{center}
\end{figure*}

The lattice graph of the system with monomers at sites $i$ and $j$
differs from the original lattice graph in that sites $i$ and $j$,
as well as all bonds connected to them, are removed. This modified
lattice graph is still planar so the Pfaffian method can again be
applied.\cite{fs-63,fms-02} Due to the bond removals, the question of
signs for the Kasteleyn matrix elements needs to be readdressed. Let us
consider the generic situation when the monomers are sufficiently far
apart that they sit on separate faces; for the star lattice each of the faces
hosting a monomer has 21 bonds along its perimeter. Consider a ``path of faces''
connecting the two monomer sites, with all but the two end faces on this path
being dodecagons. Neighboring faces along this path then share one linking bond.
It can be seen that the original choice of Kasteleyn arrows violates the clockwise-odd
sign rule around the two end faces hosting the monomers, while the sign rule is respected on the other faces on
the path. A valid choice of signs is then obtained by reversing the Kasteleyn arrows
on each of the single bonds shared by neighboring faces along the path, such that one
arrow is reversed around each of the two end faces, and two arrows are reversed
around each of the interior faces.\cite{fs-63,fms-02} An example (which we will
return to below) is shown in Fig. \ref{fig:mm-fermion}.

We now give a derivation of the general expression for $m(i,j)$ in the fermionic approach (see
also Ref. \onlinecite{fms-02}). We have $\mathcal{Z}_{mm}(i,j)/\mathcal{Z}=|Z_{mm}(i,j)/Z|$ where, by
analogy with the original lattice graph, $Z_{mm}(i,j)$ is given by
\be
Z_{mm}(i,j)=\int \mathcal{D}\psi_{i,j}\exp(-S_{i,j}).
\label{Zmm-fermi-def}
\ee
Here $\int \mathcal{D}\psi_{i,j}=\int \prod'_{l}d\psi_l$ and $S_{i,j}=\frac{1}{2}\sum_{k,l}'\psi_k \bar{A}_{kl}\psi_l$,
where $\bar{A}$ is the modified Kasteleyn matrix (i.e. with the necessary sign changes) and the prime on the
product and sum means that the Grassmann variables for sites $i$ and $j$ are excluded. To express
$Z_{mm}(i,j)$ as an expectation value with respect to the original lattice graph, we use that $\exp(-S_{i,j})$ can
be written as
\begin{eqnarray}
& & \hspace{-0.7cm}\exp\bigg(\psi_i \sum_{k} A_{ik}\psi_k\bigg)
\exp\bigg(\psi_j\sum_{l} A_{jl}\psi_l\bigg)\exp(-\psi_i A_{ij}\psi_j) \nonumber \\ & \times &
\Bigg(\prod_{(mn)\in \Lambda}(1+2\psi_m A_{mn}\psi_n)\Bigg)\exp(-S)
\label{mod-S}
\end{eqnarray}
Here the fourth factor that takes care of the sign reversals is a ``string'' of the same type encountered
in the fermionic calculation of the vison-vison correlation function in Sec. \ref{vv-fermi}; the product
again goes over the bonds whose Kasteleyn arrows are reversed.
Furthermore, $\int \mathcal{D}\psi_{i,j}=\int \mathcal{D}\psi\; \psi_i \psi_j$ (up to a possible sign difference).
The presence of $\psi_i \psi_j$ here implies that when (\ref{mod-S}) is inserted into (\ref{Zmm-fermi-def}),
the first three factors in (\ref{mod-S}) can be replaced by 1, as the omitted terms will
give no contribution to $Z_{mm}(i,j)$ owing to $\psi_i^2=\psi_j^2=0$. Thus\cite{fms-02}
\be
m(i,j) = \left|\left\langle \psi_i \Bigg(\prod_{(kl)\in \Lambda}(1+2\psi_k A_{kl}\psi_l)\Bigg)\psi_j
\right\rangle \right|.
\label{mij-fermi}
\ee

\begin{figure*}[htb]
\begin{center}
\centerline{\includegraphics[scale=0.6]{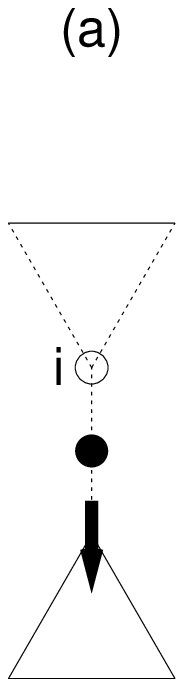}\hspace{1.2cm}
            \includegraphics[scale=0.6]{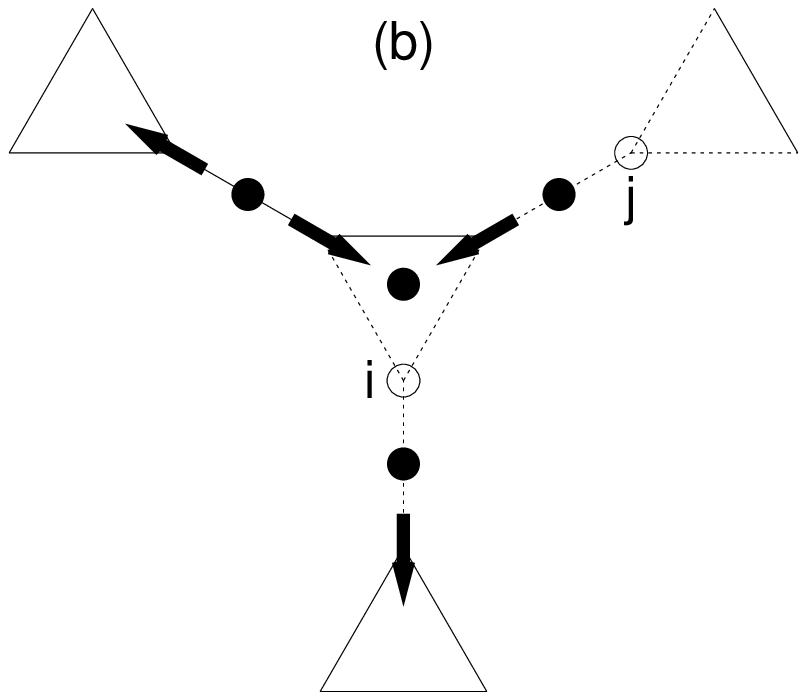}\hspace{1.2cm}
            \includegraphics[scale=0.6]{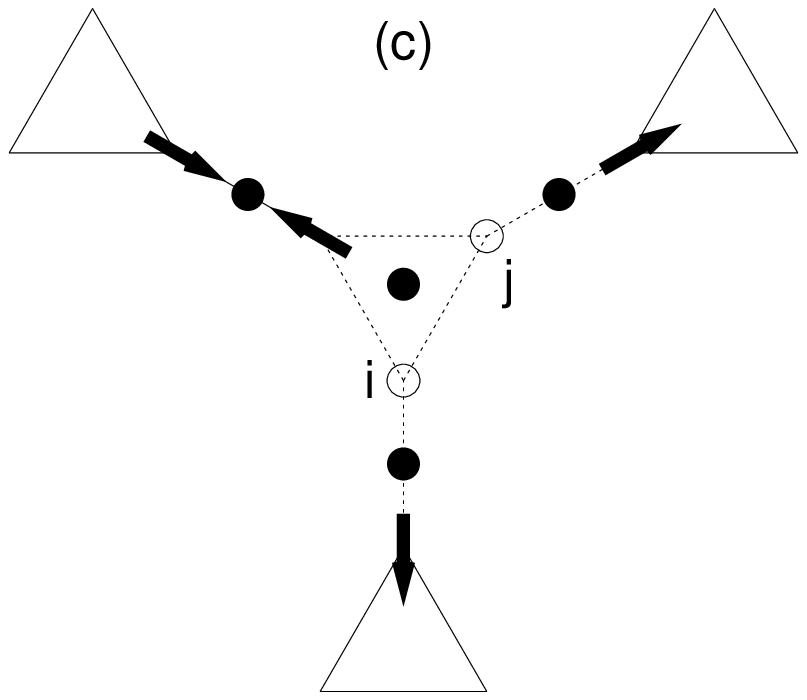}\hspace{1.2cm}
            \includegraphics[scale=0.6]{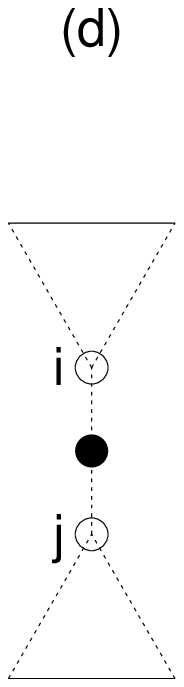}}
\caption{Arrow calculation of monomer-monomer correlation function. (a) Imprint (set of
spent arrows and constraints) for a single monomer located at site $i$. The open circle denotes
the removed arrow on the site. The filled circle denotes the spent constraint. (b), (c), and (d):
Imprints for some two-monomer configurations. For each of these, $m(i,j)$ can be found by inserting
the number of spent arrows and constraints into Eq. (\ref{rat-2}).}
\label{fig:mm-arrow}
\end{center}
\end{figure*}

We will now evaluate Eq. (\ref{mij-fermi}) for the case that the left (right) monomer is at site $i$ ($j$) where
$i=(x,y,6)$ or $(x,y+1,2)$ and $j=(x+n,y,6)$ or $(x+n,y+1,2)$, where $n>0$. Thus the $x$ coordinates of the two monomers
differ by $n$. The directions of the Kasteleyn arrows on the $n-1$ vertical linking bonds between $i$ and $j$ are
reversed. The particular case $i=(x,y,6)$, $j=(x+n,y+1,2)$ with $n=6$ is illustrated in Fig. \ref{fig:mm-fermion}.
Consider the expression
\be
\langle \psi_i\left[\prod_{\ell=1}^{n-1}(1+2\psi_{x+\ell,y,6}\psi_{x+\ell,y+1,2})\right]\psi_j\rangle
\label{mon-exp-val}
\ee
whose absolute value is $m(i,j)$. When evaluating this using Wick's theorem, all terms involving
a $1$ from one or more of the
factors $(1+2\psi_{x+\ell,y,6}\psi_{x+\ell,y+1,2})$ will give zero contribution due to the property
(\ref{G-zero}) of the Green function. Hence the only surviving term in (\ref{mon-exp-val}) is the
one with all $2n$ Grassmann variables present,
\be
\langle \psi_i\left[\prod_{\ell=1}^{n-1}2\psi_{x+\ell,y,6}\psi_{x+\ell,y+1,2}\right]\psi_j\rangle
\equiv \tilde{m}_{\alpha\beta}(n).
\ee
Here $\alpha$ and $\beta$ are the intra-cell coordinates of sites $i$ and $j$ respectively, i.e.
$\alpha$ and $\beta$ can take the values 2 and 6. We find $\tilde{m}_{62}(1)=\tilde{m}_{66}(1)=-\tilde{m}_{22}(1)
=-\tilde{m}_{26}(1)=1/4$, while for $n\geq 2$,
\begin{eqnarray}
\tilde{m}_{\alpha\beta}(n) &=& 2[\tilde{m}_{\alpha,6}(n-1)\langle \psi_{x+n-1,y+1,2}\psi_j\rangle \nonumber \\ &-&
\tilde{m}_{\alpha,2}(n-1)\langle \psi_{x+n-1,y,6}\psi_j\rangle] \nonumber \\ &=&
-\frac{1}{2}[\tilde{m}_{\alpha,6}(n-1)+\tilde{m}_{\alpha,2}(n-1)].
\end{eqnarray}
Thus $\tilde{m}_{\alpha\beta}(n)$ is independent of $\beta$.
Defining $\tilde{m}_{\alpha}(n)\equiv \tilde{m}_{\alpha\beta}(n)$, we have
\be
\tilde{m}_{\alpha}(n)=-\tilde{m}_{\alpha}(n-1) \quad \quad (n\geq 2),
\ee
with $\tilde{m}_{6}(1)=-\tilde{m}_2(1)=1/4$. It follows that the monomer-monomer correlation
function $m(i,j)=|\tilde{m}_{\alpha}(n)|=1/4$ for any $n$, including the limit $n\to\infty$.

We note that both the monomer-monomer correlation function considered here and the vison-vison correlation
function considered in Sec. \ref{vv-corr} contain a ``string'' in the fermionic formulation. This string
causes the number of pairings from Wick's theorem to grow exponentially with the vison/monomer separation.
However, in both
cases the property (\ref{G-zero}) comes to the rescue by making most of these pairings vanish. As a result
the calculations are analytically tractable regardless of the vison/monomer separation. This should be
contrasted with the situation for the triangular-lattice CDM discussed in Ref. \onlinecite{fms-02}, for
which the exponential growth of the number of pairings made the fermionic calculation of the monomer-monomer
correlation function prohibitively difficult for all but the smallest values of the monomer-monomer separation.
Instead this calculation was there carried out with a determinant formulation first used by Fisher and Stephenson for
the square lattice.\cite{fs-63} This determinant formulation was also used for the Pfaffian
calculation of the monomer-monomer correlation function of the kagome lattice CDM recently
reported in Ref. \onlinecite{lyc-08}.

Despite the essential simplifications resulting from Eq. (\ref{G-zero}), the fermionic approach to the
monomer-monomer correlation function is still less economical and less intuitive than the arrow approach
to be discussed below. For some cases of small monomer separation, the fermionic calculation of $m(i,j)$
on the star lattice also has an additional technical complication: the modified lattice graph has \textit{cut points},
\cite{cut-points} the existence of which violates the conditions for the explicit proof of the clockwise-odd rule given
in Ref. \onlinecite{k-67}. While this complication can be dealt with (the Pfaffian method can be
used also for a graph with cut points\cite{k-67}), and one reassuringly finds\cite{cut-points-calc}
the same result for $m(i,j)$ as obtained from the arrow representation, the arrow derivation is much
simpler, and therefore we do not discuss these fermionic calculations further here.

\subsection{Arrow approach}

The monomer-monomer correlation function (\ref{mm-corr-def}) is yet another example of a ratio
of the general type discussed in Sec. \ref{ratio-arrow} that can be calculated using the arrow approach. Thus we
must again identify the appropriate values of $\Delta \mathcal{N}_a$ and $\Delta \mathcal{N}_c$ in Eq. (\ref{rat-2}).
Let us start by looking at the situation when the two monomer sites $i$ and $j$ are far apart.
Consider the monomer at one of the sites, say $i$. Its contributions to $\Delta \mathcal{N}_a$ and
$\Delta \mathcal{N}_c$ are $-2$ and $-1$, respectively. Here, the two spent arrow ($Z_2$) degrees of freedom
are the (removed) arrow on $i$ and the arrow on the site at the opposite end of the (removed) linking bond that
$i$ was originally one endpoint of; in the presence of the monomer this latter arrow must point into the neighboring
triangle. Furthermore, the spent constraint contributing to $\Delta \mathcal{N}_c$ is the constraint on this
(removed) linking bond.\cite{mon-net-zero} The set of spent arrows and constraints for an individual
monomer will be referred to as the monomer \textit{imprint} (see Fig. \ref{fig:mm-arrow}(a)).
When the two monomers are far apart their contributions to $\Delta \mathcal{N}_a$ and $\Delta \mathcal{N}_c$
simply add, giving $m(i,j)=2^{2(-2-(-1))}=1/4$. This result is of course also valid as the separation
between the monomers is increased to infinity, and
is independent of the relative orientation between the underlying lattice graph and the vector connecting
the two monomers.

Next, let us analyze what happens when the two monomers are moved closer to each other. As long as the monomers
are not too close, the situation remains the same: The two monomer imprints do not overlap, nor do
they touch the same triangle, so $\Delta \mathcal{N}_a$ and $\Delta \mathcal{N}_c$ are given by the sum of the
separate contributions from each monomer. Thus $\Delta \mathcal{N}_a-\Delta \mathcal{N}_c$ is also additive,
giving $m(i,j)=1/4$. Only when the two monomers get so close that their relative positions are as shown in Fig.
\ref{fig:mm-arrow}(b) do their respective imprints touch the same triangle (the middle one in the figure). Now
$\Delta \mathcal{N}_a$ and $\Delta \mathcal{N}_c$ are no longer additive, but as the additional contributions to
$\Delta \mathcal{N}_a$ and $\Delta \mathcal{N}_c$ are equal, $\Delta \mathcal{N}_a-\Delta \mathcal{N}_c$ is still
additive, so again $m(i,j)=1/4$. This continues to hold when $j$ is moved even closer so that $i$ and $j$ are
connected by a triangle bond, as shown in Fig. \ref{fig:mm-arrow}(c). If, on the other hand, $j$ approaches $i$
from the side of $i$'s linking bond, they will finally end up being connected by this linking bond, as shown in
Fig. \ref{fig:mm-arrow}(d). In this case $\Delta \mathcal{N}_a=-2$ and $\Delta \mathcal{N}_c=-1$
so $m(i,j)=1/2$.

\subsection{Connection to spin-spin correlations in the
square-lattice Ising model}
\label{mon-Ising}

As shown by Moessner and Sondhi\cite{ms-03} (see also Ref. \onlinecite{hkms-03}),
by using Fisher's mapping\cite{f-66} between the ferromagnetic
Ising model on the square lattice and the classical dimer model on the star lattice, the spin-spin correlation
function for the spin model can be related to a sum of monomer-monomer correlation functions for the
dimer model. In our notation, this relation can be written
\be
\langle S(x,y)S(x',y')\rangle = \sum_{\alpha,\alpha'=3,4}m(x,y,\alpha;x',y',\alpha').
\label{spin-monomer}
\ee
Here $S(x,y)$ is the Ising spin located at position $(x,y)$ of the square lattice. In Fisher's mapping, the
dimer weights on the bonds corresponding to the original square lattice (see Fig. \ref{fig:star-fisher}(left))
are $1/\tanh(\beta J_{ij})$ (where $\beta$ is the inverse temperature and $J_{ij}$ is the ferromagnetic Ising
exchange parameter) while the dimer weights on the bonds internal to the unit cell
(see Fig. \ref{fig:star-fisher}(right)) are 1. The relation (\ref{spin-monomer}) implies that the phase transition
of the Ising model between states with unbroken/broken spin rotation symmetry maps onto a confinement-deconfinement
transition in the dimer model.\cite{ms-03} The equal-weight dimer model considered in this paper is
obtained in the zero-temperature limit $\beta\to\infty$ of the Ising model, in which case the spins are perfectly
ordered so the lhs of Eq. (\ref{spin-monomer}) equals 1. Let us use our results for
$m(i,j)$ to evaluate the rhs of Eq. (\ref{spin-monomer}) for this case. When $(x,y)\neq (x',y')$
all four monomer-monomer correlation functions in (\ref{spin-monomer})
equal $1/4$, so the rhs is $4\cdot 1/4=1$. On the other hand, when $(x,y)=(x',y')$ the
two monomer-monomer correlation functions with $\alpha=\alpha'$ vanish since with both monomers on the same
site, no dimer coverings are possible (as there is an odd number of sites left to form dimers).
As the two remaining terms have $m(i,j)=1/2$, the sum is now $2\cdot 0 + 2\cdot 1/2=1$. Thus our
results for $m(i,j)$ do indeed satisfy Eq. (\ref{spin-monomer}).

\section{Comparison with dimers on the kagome lattice}
\label{comp-kagome}

As briefly summarized in Sec. \ref{intro}, dimers on the kagome lattice have very interesting
properties.\cite{ez-93,mis-kag-1,mis-kag-2,wang-wu-07,wu-wang-08,wang-wu-08,lyc-08} One may
consider the existence of an arrow representation of dimer coverings on the kagome
lattice\cite{ez-93,mis-kag-1,mis-kag-2} to be a ``fundamental'' property, in the sense that the
other properties can be derived from it. Alternatively, using the Pfaffian method, the ``fundamental''
property is that the Green function of the kagome lattice vanishes beyond a very short
distance.\cite{wang-wu-07,wang-wu-08}

As shown in this paper, the properties of dimers on the star lattice closely parallel those on the kagome lattice.
Again, the most fundamental properties, from which all others can be shown to follow, are the existence
of an arrow representation, or,
alternatively, the Green function satisfying Eq. (\ref{G-zero}). (Note that the number of dimer coverings $\mathcal{Z}$
considered in Sec. \ref{ndc} is also related to the Green function: $\mathcal{Z}\sim \sqrt{\det A} =1/\sqrt{\det G}$.)
In the following we will discuss the similarities and connections between dimers on the star and kagome lattices
in some more detail. We emphasize that, unless explicitly stated otherwise, all comparisons in this section
between classical dimers on these two lattices (including those made above) are for the case when all dimers have
equal weights, which is the case we have considered in this paper. At the end we will briefly comment on the
star-lattice CDM for unequal dimer weights, whose properties are more conventional.

First, however, let us note a close relationship between the star and kagome lattices
themselves.\cite{rokhsar-90,suding-ziff,mis-sin,dusuel} These two lattices can be regarded as ``expanded'' and
``reduced'' versions of each other, in the following sense: Starting from the kagome lattice with $N_k$ sites,
touching triangles can be separated by splitting their common site into two sites connected by a new bond.
The resulting ``expanded'' lattice is a star lattice with $N_s=2N_k$ sites (the new bonds are the linking bonds
of the star lattice). Reversely, starting from a star lattice with $N_s$ sites, the linking bonds can be shrunk
to zero length and the two sites at its ends merged into one. The resulting ``reduced'' lattice is a kagome
lattice with half as many sites.

\textit{The number of dimer coverings.} For a kagome lattice of $N_k$ sites
embedded on a closed surface the number of dimer coverings
is\cite{ndc-kag,ez-93,mis-kag-1,mis-kag-2,wu-wang-08} $\mathcal{Z}_k(N_k)=2^{N_k/3+1}$.
As shown in Sec. \ref{ndc-arrows}, the corresponding result for a star lattice with $N_s$ sites
is $\mathcal{Z}_s(N_s)=2^{N_s/6+1}$. Thus the kagome and star lattices that are each other's reduced/expanded
lattices with $N_s=2N_k$ have exactly the same number of dimer coverings:
\be
\mathcal{Z}_s(N_s)=\mathcal{Z}_k(N_k).
\label{ndc-star-kagome}
\ee
A related special property of the kagome and star lattice is that they have very simple expressions for
the entropy per site/dimer (a simple rational number times $\log 2$), in contrast to e.g. the square,
honeycomb and triangular lattice.\cite{wu-review} The squagome\cite{mis-kag-1,mis-kag-2} and triangular-kagome
lattice\cite{lyc-08} also have this property; we will comment more on this in Sec. \ref{reduced}.

A consequence of Eq. (\ref{ndc-star-kagome}) is that the Hilbert spaces of QDMs defined on the kagome/star lattices
that are reduced/expanded versions of each other have the same dimension. One can therefore construct one-to-one
mappings between the basis states (dimer coverings) in these Hilbert spaces. One
possible mapping is as follows: Pick an arbitrary dimer covering $|c_k\rangle$ on the
kagome lattice and map this to an arbitrary dimer covering $|c_s\rangle$
on the star lattice. Then map the dimer covering $\prod \hat{\sigma}^x(H)|c_k\rangle$ on the kagome lattice to
the dimer covering $\prod \hat{\sigma}^x(D)|c_s\rangle$. Here the product runs over some set of hexagons in the
kagome case and over the associated set of dodecagons in the star lattice case (a given hexagon in the kagome lattice
is associated with a unique dodecagon in the star lattice by the ``extension/reduction'' procedure described
earlier). In this way one gets a one-to-one mapping between dimer coverings in given topological sectors
on the two lattices. This mapping is particularly ``natural'' from the point of view of the QDM Hamiltonian
(\ref{qdm-ham}) and its kagome analogue considered in Ref. \onlinecite{mis-kag-1}, as they are given by a sum of
$\hat{\sigma}^x$ operators on all hexagons/dodecagons.

\textit{Dimer-dimer correlations.} On both the kagome and star lattice a dimer is uncorrelated with any dimer
further away than on a neighboring triangle. For the kagome lattice this means that a dimer is correlated
with dimers that are up to three bonds away,\cite{mis-kag-1,wang-wu-07,wang-wu-08} while for the star lattice,
as shown in Sec. \ref{d-corr}, dimers
on triangle bonds are correlated with dimers up to four bonds away, while dimers on linking bonds are correlated
with dimers that are only up to two bonds away (in fact, only with the triangle-bond subset of these,
as a linking bond dimer is uncorrelated even with dimers on the nearest linking bonds that are only two bonds away).
Furthermore, comparing Fig. \ref{fig:dd-corr}(a) with the corresponding figure for the kagome lattice (the latter can
be constructed from the results in Ref. \onlinecite{wang-wu-08}, or calculated from the arrow representation), one finds
that the numbers on the triangle in the middle are identical while those on the three outer triangles have the same
magnitude but opposite sign (the linking bonds are of course not there in the kagome lattice figure).

\textit{Vison-vison correlations.} The vison-vison correlation function $v(I,J)$ vanishes for both lattices
(here $I$ and $J$ are hexagon/dodecagon sites, respectively). In both cases this is most easily seen from the
fact that, as discussed in Sec. \ref{vv-corr}, the state with visons at $I$ and $J$ is an eigenstate of the
QDM Hamiltonian, and the absolute value of the overlap between this two-vison excited state and the ground state
is just the vison-vison correlation function.

\textit{Monomer-monomer correlations.} The monomer-monomer correlation function $m(i,j)$ on the kagome lattice
equals $1/4$ for all monomer sites $i$ and $j$.\cite{mis-lhu,lyc-08} As shown in Sec. \ref{mm-corr}, the same
result holds for the star lattice, with one exception: when $i$ and $j$ are connected by a linking bond, $m(i,j)=1/2$.

Finally, we note that the result (\ref{G-zero}) for the star-lattice Green function is
limited to the case of equal dimer weights that we have considered in this paper. As a consequence, the very
special properties of the star-lattice CDM for equal dimer weights do not carry over to unequal dimer weights.
In contrast, the special properties of dimers on the kagome lattice are more robust, as the result (\ref{G-zero})
continues to hold also if the dimer weight is allowed to depend on the orientation of the
bond.\cite{wang-wu-07,wang-wu-08}

\section{General Fisher lattices and their ``reduced'' lattices}
\label{gen-fisher}

It is interesting to ask whether the results obtained here for the star lattice can be generalized to other
lattices. A natural class of lattices to consider in this respect are those hosting the dimers in Fisher's
mapping from an Ising model to a dimer problem,\cite{f-66} of which the star lattice is one particular example.
Any such ``terminal'' lattice in Fisher's mapping will here be referred to as a general Fisher lattice, or
just Fisher lattice for short. The sites (vertices) in a general Fisher lattice can have coordination number
(henceforth called degree) 1, 2, or 3. Sites of degree 3 are part of a triangle of three sites connected by
bonds which we call triangle bonds. All other bonds in the Fisher lattice will be called linking bonds.

\subsection{Arrow representation and the number of dimer coverings for general Fisher lattices}

An arrow representation of dimer coverings exists also for a general Fisher lattice. Again,
the arrows are Ising degrees of freedom living on the lattice sites, and each triangle and each linking bond contribute
an arrow constraint. Using the arrow representation, one can calculate various properties of dimers on general Fisher
lattices, analogously to what we have done for the star lattice. As an example, we will show that the arrow
representation gives the correct answer for the number of dimer coverings on an arbitrary Fisher lattice. We do
this by first using Fisher's mapping to calculate this number, and then we show that the arrow representation
gives the same result.

Thus let us consider Eq. (11) in Ref. \onlinecite{f-66}, which relates the Ising model partition function
(defined in Fisher's Eq. (3)) to the generating function of dimer coverings on the associated Fisher lattice.
We are interested in the case of equal dimer weights,
i.e. $v_{ij}=1$, for which this generating function reduces to the number of dimer coverings $\mathcal{Z}$ on the
Fisher lattice. This gives (in our notation) $\mathcal{Z}_I = 2^{N_I - N_{I,b}}\mathcal{Z}$.
Here $\mathcal{Z}_I$ is the partition function of the Ising model and $N_I$ and $N_{I,b}$ are
the number of sites and bonds in the lattice on which this Ising model is defined (which we will refer to as the
Ising lattice). As the equal-weight case maps to the zero-temperature limit of the Ising model, $\mathcal{Z}_I=2$
since only the two perfectly ordered states related by time reversal contribute to $\mathcal{Z}_I$ in this limit.
Thus
\be
\mathcal{Z}=2^{N_{I,b}-N_I+1}.
\label{z-gen-fisher}
\ee
Next, let us calculate $\mathcal{Z}$ using the arrow representation. As in Sec. \ref{ndc-arrows}, we have
$\mathcal{Z}=2^{\mathcal{N}_a-\mathcal{N}_c}$ where $\mathcal{N}_a$ is the number of $Z_2$ arrow degrees of
freedom and $\mathcal{N}_c$ is the number of independent arrow constraints. For a general Fisher lattice we
have $\mathcal{N}_a=N_F-N_F^{(1)}$ where $N_F$ is the
number of sites on the Fisher lattice and $N_F^{(1)}$ is the number of sites of degree 1. The reason for
subtracting $N_F^{(1)}$ is that an arrow on a site of degree 1 has to point towards the single bond
connected to the site (because the site has to be touched by a dimer) and is therefore not a $Z_2$
degree of freedom. The number of independent arrow constraints is $\mathcal{N}_c=N_T+N_l-1$ where $N_T$ is the number
of triangles and $N_l$ is the number of linking bonds in the Fisher lattice. As in Sec. \ref{ndc-arrows}
the term $-1$ in $\mathcal{N}_c$ comes about because one of the $N_T+N_l$ constraints can be deduced from the
others. We show in Appendix \ref{proof-eqs} that
\be
(N_F-N_F^{(1)}) - (N_T + N_l) = N_{I,b}-N_I
\label{eq-1}
\ee
which means that the result (\ref{z-gen-fisher}) is indeed reproduced by the arrow representation. In Appendix
\ref{proof-eqs} we also show that
\be
N_{I,b}-N_I = \frac{1}{6}(N_F-N_F^{(2)})-\frac{2}{3}N_F^{(1)}.
\label{eq-2}
\ee
For a general Fisher lattice we therefore have
\be
\mathcal{Z}=2^{(N_F -N_{F}^{(2)})/6 - 2N_{F}^{(1)}/3+1}.
\label{Z-gen-fisher-final}
\ee
For Fisher lattices with no sites of degree 1 or 2 this simplifies to
\be
\mathcal{Z}=2^{N_F/6+1}.
\label{Z-simp-fisher}
\ee
This simplification applies e.g. to the star lattice graph considered in Sec. \ref{ndc-arrows}, as all
its sites have degree 3; the result for $\mathcal{Z}$ is indeed the same as found there. In Appendix
\ref{check-fisher} we check the result (\ref{Z-gen-fisher-final}) for some Fisher lattices with sites of
degree 1 and/or 2.

\subsection{The ``reduced'' lattice of a Fisher lattice}
\label{reduced}

The kagome lattice is the ``reduced'' lattice of the star lattice in the sense described in Sec. \ref{comp-kagome}.
An analogous reduction procedure, consisting of shrinking the linking bonds to zero and merging all
sites that come together in this process, can be carried out for any Fisher lattice F. Unless F is trivial by not
containing any triangles, the sites in the reduced lattice R so obtained have coordination number four.

As a first example, consider the Fisher lattice associated with the triangular-lattice Ising model, which can be
shown to consist of triangles, octagons, and ``16-gons.'' The reduced Fisher lattice R is a squagome
lattice.\cite{siddharthan,mis-kag-2} This is a lattice of corner-sharing triangles and therefore has an arrow
representation of its own as noted in Refs. \onlinecite{mis-kag-1,mis-kag-2}.

As a second example, consider the kagome-lattice Ising model. The lattice E (this notation is
explained in Appendix \ref{app-fisher}) is then the star lattice. The Fisher
lattice F is obtained by replacing each site of the star lattice with a triplet of sites connected by
a triangle of additional bonds. The reduced lattice R of this Fisher lattice is the triangular-kagome
lattice recently considered in Ref. \onlinecite{lyc-08}. Although this is not a lattice of corner-sharing
triangles (as its triangles also share edges), it still has an arrow representation. To see this,
note that any dimer on this lattice can be uniquely associated with an ``outer small'' triangle.
(Each ``big'' triangle consists of 4 small triangles: 3 outer and 1 inner). Therefore an arrow representation
exists in which each arrow must point into one of the two outer small triangles that its site connects,
i.e. the arrows are again $Z_2$ variables. The fact that the inner small triangles have a different status
than the outer small ones is reflected in the fact that the former triangles are not present in the
Fisher lattice F but only appear after F has been reduced to R.

The number of dimer coverings on F and R are closely related, as we have already seen for the star and kagome
lattice in Sec. \ref{comp-kagome}. The reduction of F to R involves
removing one site for each linking bond, and therefore the difference $\mathcal{N}_a-\mathcal{N}_c$ is the same
in F and R, and thus so is the number of dimer coverings (cf. Eq. (\ref{ndc-star-kagome})). Since the number
of sites on the two lattices are related by $N_F=2N_R$, the entropy per site on R in the thermodynamic limit
is, from Eq. (\ref{Z-simp-fisher}), given by $s_R=(1/3)\log 2$, which agrees with previous calculations
in the literature.\cite{ndc-kag,ez-93,mis-kag-1,mis-kag-2,wu-wang-08,lyc-08} The fact that the entropy per
site is the same for the kagome, squagome, and triangular
kagome lattice can thus be understood to be a consequence of the fact that each of these lattices is the reduced
lattice of a Fisher lattice for which Eq. (\ref{Z-simp-fisher}) holds.

Based on the existence of the arrow representation as well as recent studies of what we have here dubbed reduced Fisher
lattices,\cite{mis-kag-1,mis-kag-2,lyc-08} we expect that also other properties of dimers on general Fisher lattices
and their reduced lattices should in general be closely related and resemble those of the star and kagome lattice
(one exception is the purely one-dimensional Fisher lattices considered in Sec. \ref{check-fisher}).

\begin{figure*}[htb]
\begin{center}
\centerline{\includegraphics[scale=0.8]{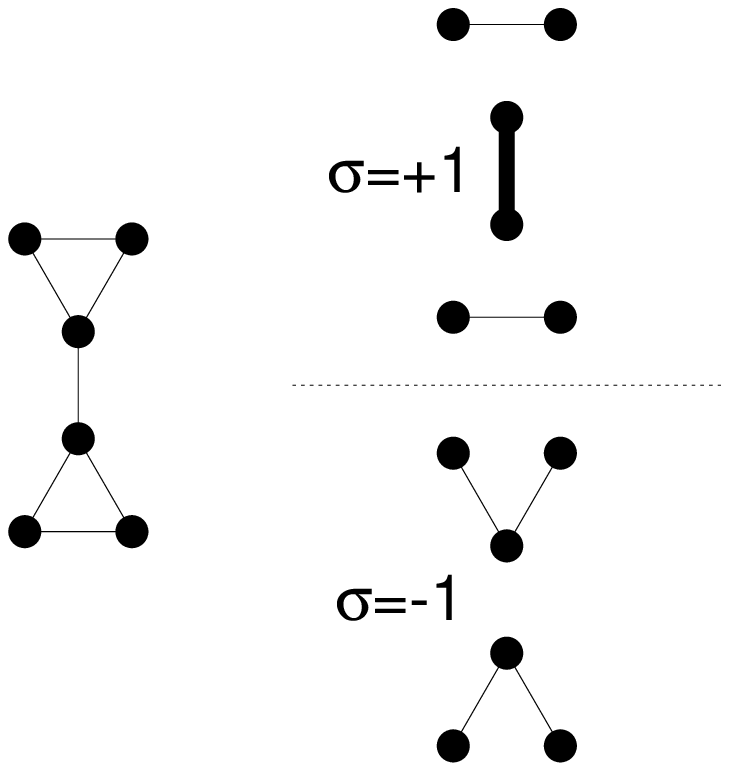}\hspace{2cm}
            \includegraphics[scale=0.8]{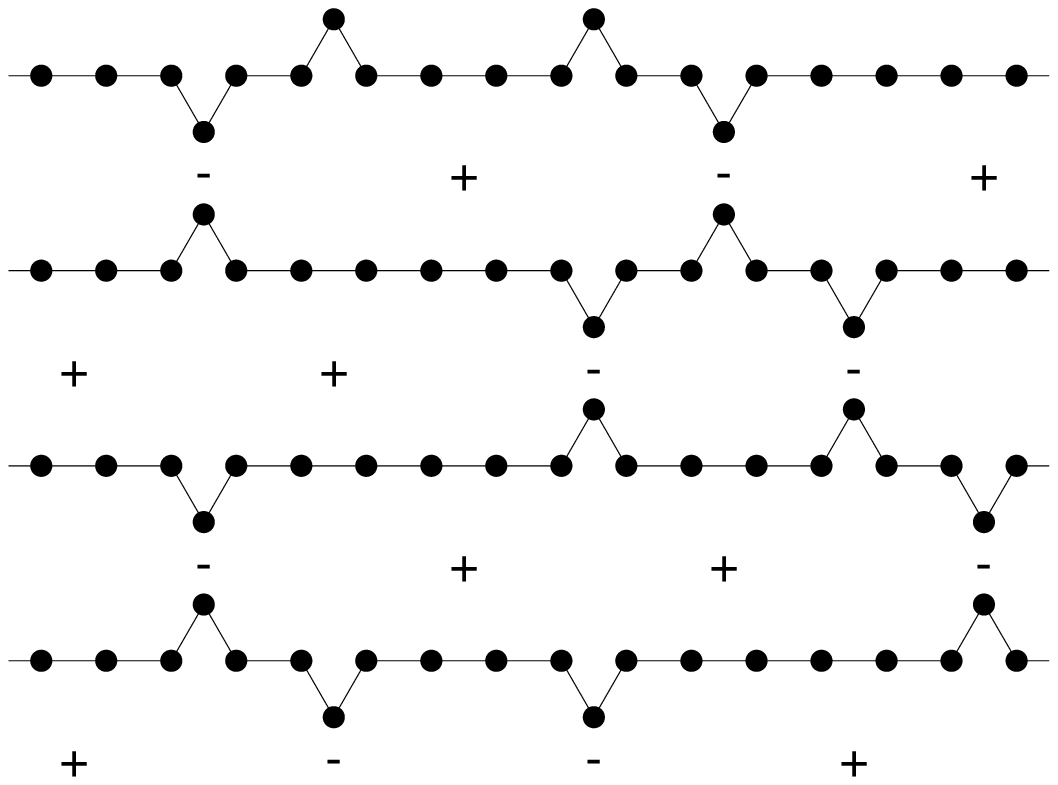}}
\caption{In this figure filled circles represent lattice sites.
Left: 6-site unit cell for the star lattice. An Ising pseudospin $\sigma=\pm 1$ is associated
with the vertical linking bond connecting the two triangles. Middle: The unit cell graph that results
from deleting the bonds that, for the given value of $\sigma$, cannot be occupied by a dimer. (The thick line
in the $\sigma=+1$ graph is the dimer on the linking bond.) Right: The decoupled horizontal
loops resulting from deleting the appropriate bonds in all unit cells for a particular pseudo-spin
configuration (the pseudo-spins are shown as signs $\pm$) in a system with $L=4$. (For ease of drawing,
the vertical dimers in the $\sigma=+1$ unit cells are not shown and the bonds connecting different
unit cells have all been made horizontal.)}
\label{fig:dc-ps}
\end{center}
\end{figure*}

\section{Final remarks}
\label{conc}

We conclude by mentioning another lattice recently considered in the literature that also has an arrow
representation. This lattice, which is a hybrid of the kagome and star lattices, is shown in Fig. 1 (right)
in Ref. \onlinecite{ms-03}, which considered a dimer problem on this lattice. Using the arrow representation
one can e.g. reproduce the result for equal-weight dimers ($u=1$) obtained in Ref. \onlinecite{ms-03} with the
Pfaffian method.

\acknowledgments

We thank Gr\'{e}goire Misguich and Roderich Moessner for comments on the manuscript. This research was
supported by the Australian Research Council.

\appendix

\section{An alternative pseudo-spin representation of dimer coverings on the star lattice}
\label{dc-gen}

Recently Dhar and Chandra introduced an alternative pseudo-spin representation of dimer coverings on
lattices of corner-sharing triangles.\cite{dc} Wu and Wang used this representation to
give a derivation of the number of dimer coverings on a finite kagome lattice.\cite{wu-wang-08}
In this Appendix we generalize this pseudo-spin representation to the star lattice and use it
to give an alternative simple derivation of the result (\ref{Zt-result}) for the
number of dimer coverings on a finite star lattice on a torus.

Consider the star lattice as shown in Fig. \ref{fig:star-lattice}. Here we will choose the
6-site unit cells to be oriented as shown in Fig. \ref{fig:dc-ps} (left panel). A unit cell will be labeled
$(k,\ell)$ where $k=1,\ldots,K$ and $\ell=1,\ldots,L$. Here $L$ is the number of horizontal
rows of these unit cells and $K$ is the number of unit cells in each such row. We will consider
the star lattice graph to have a torus geometry by connecting opposite ends in the horizontal and vertical
direction. The number of unit cells is $N/6$ where $N=6KL$ is the total number of sites in the lattice.

We now associate an Ising pseudo-spin variable $\sigma_{k\ell}=\pm 1$ with the vertical linking bond in
unit cell $(k,\ell)$. We define $\sigma_{k\ell}=+1$ ($-1$) if this linking bond is occupied (not occupied)
by a dimer. Let us consider how many dimer coverings are associated with a given pseudo-spin configuration
$\{\sigma_{k\ell}\}$. One sees that for a given value of $\sigma_{k\ell}$ some bonds in the unit cell
$(k,\ell)$ cannot be occupied by a dimer. By removing these bonds in all the unit cells one obtains a new
lattice graph that consists of $L$ disconnected horizontal closed loops of bonds. This is illustrated
in Figs. \ref{fig:dc-ps} (middle and right panels).
If all of these disconnected loops contain an even number of sites, the
number of dimer coverings associated with the pseudo-spin configuration is $2^{L}$, since there are two dimer
coverings for each loop. On the other hand, if some loops have an odd number of sites, no dimer coverings
are possible for the given pseudo-spin configuration.

It remains to find the number of pseudo-spin configurations that are consistent with dimer coverings. To this end,
we define the variables\cite{wu-wang-08}
\be
\tau_{\ell}=\prod_{k=1}^{K}\sigma_{k\ell}, \quad \ell=1,\ldots,L.
\ee
If $\sigma_{k\ell}=+1$ the unit cell $(k,\ell)$ contributes 2 sites to each of the two loops above and below it,
while if $\sigma_{k\ell}=-1$ the unit cell contributes 3 sites to each of these loops. In order for the loop
going between unit cell rows $\ell$ and $\ell+1$ to contain an even number of sites, an even number of the unit
cells contributing sites to it must have pseudo-spin $\sigma=-1$. This constraint can be written
\be
\tau_{\ell}\tau_{\ell+1}=1.
\ee
There are $L$ such constraints, one for each loop. However, only $L-1$ of these are independent, since
due to $\sigma_{k\ell}^2=1$ we have
\be
(\tau_1 \tau_2)(\tau_2 \tau_3)\ldots (\tau_{L}\tau_1)=1
\ee
so one of the constraints can be deduced from the others. The number of pseudo-spin configurations that are
consistent with dimer coverings is therefore $2^{N/6-(L-1)}$. Since each of these configurations is associated
with $2^{L}$ dimer coverings, the total number of dimer coverings is
\be
\mathcal{Z}=2^{N/6-(L-1)}\cdot 2^{L}=2^{N/6+1}
\ee
in agreement with Eq. (\ref{Zt-result}).

\section{Details for Sec. \ref{gen-fisher}}
\label{app-fisher}

\subsection{Derivation of Eqs. (\ref{eq-1})-(\ref{eq-2})}
\label{proof-eqs}

Consider an Ising lattice I composed of $N_{I}^{(q)}$ sites of degree $q$, where $q$ runs over the positive
integers. The total number of sites and bonds in this lattice are, respectively,
\begin{eqnarray}
N_I &=& \sum_{q\geq 1}N_{I}^{(q)},\\
N_{I,b} &=& \frac{1}{2}\sum_{q\geq 1}q N_{I}^{(q)}.
\end{eqnarray}
Fisher\cite{f-66} introduces an ``expanded'' lattice E in which each vertex of degree $q\geq 4$ in I is
replaced by a ``cee'' of $q-2$ vertices of degree 3 and $q-3$ additional bonds connecting these.
Thus the number of sites and bonds in E are, respectively,
\begin{eqnarray}
N_E &=& \sum_{q=1}^3 N_I^{(q)} + \sum_{q\geq 4}(q-2)N_I^{(q)},\\
N_{E,b} &=& N_{I,b}+\sum_{q\geq 4}(q-3)N_I^{(q)} = \frac{1}{2}\sum_{q=1}^3 q N_E^{(q)},
\end{eqnarray}
where in the last expression $N_{E}^{(1)}=N_{I}^{(1)}$, $N_{E}^{(2)}=N_{I}^{(2)}$, and
$N_{E}^{(3)}=N_{I}^{(3)}+\sum_{q\geq 4}(q-2)N_I^{(q)}$ (by construction, $N_E^{(q)}=0$ for $q\geq 4$).

Next, Fisher introduces a ``terminal'' lattice, which we will call the Fisher lattice F. It is obtained from E
by leaving sites of degree 1 as they are, replacing each site of degree 2 by a pair of new sites
(connected by an additional bond), and replacing each site of degree 3 by a triplet of new sites (connected by
a triangle of additional bonds). The number of sites in F is therefore
\be
N_F = N_{E}^{(1)}+2N_{E}^{(2)}+3N_{E}^{(3)}=2N_{I,b}+\sum_{q\geq 4}(2q-6)N_I^{(q)}.
\ee
Furthermore, the number of bonds in F is
\be
N_{F,b}=N_{E,b}+N_{E}^{(2)}+3N_{E}^{(3)}
\ee
since each site of degree 2 in E gives one new bond and each site of degree 3 in E gives 3 new bonds (the triangle
bonds). The number of triangles in F is
\be
N_T = N_{E}^{(3)} = N_{I}^{(3)}+\sum_{q\geq 4}(q-2)N_I^{(q)}
\ee
The number of linking bonds in F is the sum of the first two terms in $N_{F,b}$, i.e.
\be
N_l = N_{E,b}+N_{E}^{(2)} = N_{I,b}+\sum_{q\geq 4}(q-3)N_I^{(q)} + N_{I}^{(2)}.
\ee
Eqs. (\ref{eq-1}) and (\ref{eq-2}) now follow from these expressions and the fact that $N_F^{(1)}=N_I^{(1)}$ and
$N_F^{(2)}=2N_I^{(2)}$.

\subsection{Check of Eq. (\ref{Z-gen-fisher-final}) for Fisher lattices with sites of degree 1 and/or 2}
\label{check-fisher}

As a check of Eq. (\ref{Z-gen-fisher-final}) when $N_{F}^{(1)}$ and/or $N_{F}^{(2)}$ are nonzero, let us consider an
Ising model in one dimension. First consider the case of periodic boundary conditions. All sites in the Ising lattice I
are then of degree 2. The expanded lattice E is therefore identical to I. Continuing to the Fisher lattice F,
each site in E is replaced by 2 sites in F. Thus F is simply a 1D chain with an even number of sites with
periodic boundary conditions. Obviously, 2 dimer coverings are possible on F in this case, and this indeed also
follows from Eq. (\ref{Z-gen-fisher-final}) when we insert $N_F=N_{F}^{(2)}$ and $N_{F}^{(1)}=0$.

Next, consider a 1D Ising lattice with open boundary conditions. Now the two end sites in I have degree 1 while
all others have degree 2 as before. Again E is identical to I, while F has twice as many degree-2 sites (so
the total number of sites in F is even). Obviously, only 1 dimer covering is possible on F in this case, and
this is also what Eq. (\ref{Z-gen-fisher-final}) gives upon inserting $N_F=N_F^{(1)}+N_F^{(2)}$ and $N_{F}^{(1)}=2$.


\begin{thebibliography}{99}

\bibitem{k-67} P. W. Kasteleyn, in \textit{Graph Theory and Theoretical Physics},
ed. F. Harvey (Academic Press, 1967).

\bibitem{fms-02} P. Fendley, R. Moessner, and S. L. Sondhi, Phys. Rev. B \textbf{66}, 214513 (2002).

\bibitem{wu-review} F. Y. Wu, Int. J. Mod. Phys. B \textbf{20}, 5357
(2006).

\bibitem{dijkgraaf} R. Dijkgraaf, D. Orlando, and S. Reffert,
arXiv:0705.1645.

\bibitem{math} There is also a considerable mathematical literature on problems involving classical
dimers; see e.g. J. Propp, math/9904150; R. Kenyon, math/0310326.

\bibitem{k-61} P. W. Kasteleyn, Physica \textbf{27}, 1209
(1961).

\bibitem{tf-61} H. N. V. Temperley and M. E. Fisher, Phil. Mag. \textbf{6},
1061 (1961); M. E. Fisher, Phys. Rev. \textbf{124}, 1664 (1961).

\bibitem{fs-63} M. E. Fisher and J. Stephenson, Phys. Rev.
\textbf{132}, 1411 (1963).

\bibitem{k-63} P. W. Kasteleyn, J. Math. Phys. \textbf{4},
287 (1963).

\bibitem{f-66} M. E. Fisher, J. Math. Phys. \textbf{7}, 1776
(1966).

\bibitem{ms-03} R. Moessner and S. L. Sondhi, Phys.
Rev. B \textbf{68}, 054405 (2003).

\bibitem{name} The star lattice is also known under many other names, including the 3-12 or $3\cdot 12^2$
lattice,\cite{suding-ziff,wu-review} the Fisher lattice,\cite{ms-03,hkms-03} the decorated hexagonal\cite{rokhsar-90}
or expanded kagome lattice,\cite{rokhsar-90,mis-sin} and the triangle-honeycomb lattice.\cite{yao-kivelson,dusuel}
The name star lattice, which will be used in this paper, was coined in Ref. \onlinecite{suding-ziff} and was
also used in Refs. \onlinecite{richter-review}, \onlinecite{richter-2}, \onlinecite{star-exp} and (partly)
\onlinecite{mis-sin}.

\bibitem{r-et-al-rev} K. S. Raman, E. Fradkin, R. Moessner,
S. Papanikolaou, and S. L. Sondhi, arXiv:0809.3050.

\bibitem{mr-rev} R. Moessner and K. S. Raman, arXiv:0809.3051.

\bibitem{and-faz} P. W. Anderson, Mater. Res. Bull. \textbf{8}, 153
(1973); P. Fazekas and P. W. Anderson, Phil. Mag. \textbf{30}, 423
(1974).

\bibitem{wen-book} X.-G. Wen, \textit{Quantum Field Theory of Many-Body Systems}
(Oxford University Press, 2004).

\bibitem{rk} D. S. Rokhsar and S. A. Kivelson, Phys.
Rev. Lett. \textbf{61}, 2376 (1988).

\bibitem{ms-rvb} R. Moessner and S. L. Sondhi, Phys. Rev.
Lett. \textbf{86}, 1881 (2001).

\bibitem{mis-kag-1} G. Misguich, D. Serban, and V.
Pasquier, Phys. Rev. Lett. \textbf{89}, 137202 (2002).

\bibitem{flippable} A flippable loop is an even-length loop in which the bonds along the
loop alternate between being occupied and unoccupied by a dimer. The ``flip'' interchanges occupied
and unoccupied bonds and is equivalent to shifting the dimers by one bond length along the loop.

\bibitem{ez-93} V. Elser and C. Zeng, Phys. Rev. B \textbf{48}, 13647 (1993).

\bibitem{mis-kag-2} G. Misguich, D. Serban, and V. Pasquier,
Phys. Rev. B \textbf{67}, 214413 (2003).

\bibitem{wang-wu-07} F. Wang and F. Y. Wu, Phys. Rev. E \textbf{75}, 040105(R)
(2007).

\bibitem{wang-wu-08} F. Wang and F. Y. Wu, Physica A \textbf{387}, 4157
(2008).

\bibitem{mis-lhu} G. Misguich and C. Lhuillier, \textit{Two-dimensional quantum antiferromagnets}, in
\textit{Frustrated spin systems}, edited by H. T. Diep (World Scientific, 2003).

\bibitem{lyc-08} Y. L. Loh, D.-X. Yao, and E. W. Carlson, arXiv:0803.0742.

\bibitem{arch} In addition to the star lattice, the Archimedean tilings also include the square, triangular,
honeycomb, kagome, and Shastry-Sutherland lattice, as well as five other, less frequently encountered lattices;
see e.g. Refs. \onlinecite{richter-review} and \onlinecite{suding-ziff}.

\bibitem{richter-review} J. Richter, J. Schulenburg, and A. Honecker, Lect. Notes Phys.
\textbf{645}, 85 (2004).

\bibitem{richter-2} J. Richter, J. Schulenburg, A. Honecker, and D. Schmalfuss, Phys. Rev. B \textbf{70},
174454 (2004).

\bibitem{mis-sin} G. Misguich and P. Sindzingre, J. Phys.: Condens. Matter \textbf{19}, 145202 (2007).

\bibitem{star-exp} Y.-Z. Zheng, M.-L. Tong, W. Xue, W.-X. Zhang,
X.-M. Chen, F. Grandjean, and G. J. Long, Angew. Chem. Int. Ed.
\textbf{46}, 6076 (2007).

\bibitem{yao-kivelson} H. Yao and S. A. Kivelson, Phys. Rev. Lett.
\textbf{99}, 247203 (2007).

\bibitem{samuel-1} S. Samuel, J. Math. Phys. \textbf{21}, 2806 (1980).

\bibitem{hkms-03} D. A. Huse, W. Krauth, R. Moessner, and S. L. Sondhi,
Phys. Rev. Lett. \textbf{91}, 167004 (2003).

\bibitem{rokhsar-90} D. S. Rokhsar, Phys. Rev. B \textbf{42}, 2526 (1990).

\bibitem{dusuel} S. Dusuel, K. P. Schmidt, J. Vidal, and R. L.
Zaffino, arXiv:0804.4775.

\bibitem{suding-ziff} P. N. Suding and R. M. Ziff, Phys. Rev. E \textbf{60}, 275 (1999).

\bibitem{mis-sin-men} After this work was completed, we became aware that the existence
of an Ising pseudo-spin representation for dimer coverings on the star lattice, and the result
(\ref{Zt-result}) for the torus, were mentioned (without elaboration) in Ref. \onlinecite{mis-sin}.
We thank Gr\'{e}goire Misguich for drawing our attention to the discussion of the star lattice
in this paper.

\bibitem{top-loop} A loop is topologically trivial (nontrivial) if it can (can not)
be shrunk to a point. For example, on a torus, loops that wind around the system
in one or more of the two toroidal directions are topologically nontrivial.

\bibitem{consistent} To check that this assignment of pseudo-spins to the dodecagons does not produce any
contradictions, one can consider arbitrary closed walks visiting dodecagon sites of the dual lattice.
An even number of domain walls must be crossed on any such closed walk as otherwise the sign of $\sigma^z$
would have changed upon returning to the dodecagon one started from. This requirement is indeed satisfied
because all domain walls are topologically trivial loops since the two DCs are by assumption in the same
topological sector.\cite{mis-kag-2}

\bibitem{gen-comm} This generalization can be done by introducing a ``path of dodecagons'' $D_i$, $i=1,\ldots,n$
going between the two dodecagons, and using $\hat{\sigma}^z(D_i)^2=I$. For example, if $D'$ is not a neighbor of $D$
one can write $\hat{\sigma}^z(D)\hat{\sigma}^z(D')=(\hat{\sigma}^z(D)\hat{\sigma}^z(D_1))(\hat{\sigma}^z(D_1)
\hat{\sigma}^z(D_2))\ldots (\hat{\sigma}^z(D_n)\hat{\sigma}^z(D'))$. Now each pair involves nearest-neighbor
dodecagons, so the previous results can be invoked. The end result is the generalization
stated in the text.

\bibitem{top-order} X.-G. Wen and Q. Niu, Phys. Rev. B \textbf{41}, 9377 (1990).

\bibitem{dc} D. Dhar and S. Chandra, Phys. Rev. Lett. \textbf{100},
120602 (2008).

\bibitem{eliminate} To see this, choose one of the arrows in a constraint to be the `output'
and the other arrow(s) to be the `input.' All inputs are valid and determine, via the
constraint, the output. Thus a constraint can be used to eliminate the output spin
while not restricting the input spin(s). In contrast, the latter property does not hold for (e.g.) dimers on
the square lattice, for which the presence/absence of a dimer on a bond corresponds to
a bond spin $\sigma=\pm 1$, i.e. each site has a hard-core constraint $\sum_{i=1}^4 \sigma_i=-2$.
Here an input in which more than one of the three input spins is $+1$ is not valid.
Thus the constraint not only determines the output but also puts restrictions on the
input. This is consistent with the number of dimer coverings which is asymptotically
$2^{NG/(\pi \ln 2)}\approx 2^{0.42 N}$, since writing this as $2^{\mathcal{N}_a-\mathcal{N}_e}$
where $\mathcal{N}_a=2N$ is the number of spins and $\mathcal{N}_e$ is the number of eliminated
spins due to the constraints, we get $\mathcal{N}_e\approx 1.58 N$, i.e. each constraint
effectively eliminates $1.58>1$ Ising variables.

\bibitem{per-dimer} Note that the intensive entropy $(1/3)\log 2$ quoted for the star lattice
in Ref. \onlinecite{wu-review} is the entropy per dimer, not per lattice site.

\bibitem{general} More generally, the magnitude of $A_{ij}$ is set
equal to the (nonnegative) weight of the dimer that can occupy bond
$ij$; different bonds can be assigned different weights. In this paper
we will only be concerned with the case when all dimers have the
same weight, which can then be set equal to 1.

\bibitem{arrow-confuse} These fixed bond arrows (which we will
sometimes refer to as Kasteleyn arrows) should not be confused with the arrows
introduced in Sec. \ref{arrow}, which are dynamical degrees of freedom living
on the lattice sites.

\bibitem{detailed} See Ref. \onlinecite{k-67} for a detailed discussion.

\bibitem{wu-wang-08} F. Y. Wu and F. Wang, Physica A \textbf{387}, 4148
(2008).

\bibitem{iif-02} A. Ioselevich, D. A. Ivanov, and M. V. Feigelman,
Phys. Rev. B \textbf{66}, 174405 (2002).

\bibitem{unit-weight-form} If the dimer weights are not all equal to 1,
the fermionic correlation function on the right-hand side will be multiplied
by a weight-dependent prefactor; see Ref. \onlinecite{wang-wu-08}.

\bibitem{additional-ac} This also includes the case when the two non-overlapping imprints
are part of the same triangle, by each containing an
arrow of the triangle. Then the direction of the arrow on the third site of the triangle also
becomes fixed, and in turn the arrow on the other side of its linking bond. Furthermore, the
triangle constraint and the linking bond constraint are spent. Thus in this case there are
further contributions to $\Delta \mathcal{N}_a$ and $\Delta \mathcal{N}_c$ in addition to those
from the individual dimer imprints. However, since the numbers of additional spent arrows and
constraints are equal, they cancel in $\Delta \mathcal{N}_a - \Delta \mathcal{N}_c$.

\bibitem{misguich-mila-08} G. Misguich and F. Mila, Phys. Rev. B
\textbf{77}, 134421 (2008).

\bibitem{dimer-monomer} For the system with monomers, the ``normal'' sites (i.e.
those that are not occupied by a monomer) should be touched by exactly one dimer
as usual. Only the case with an even number of monomer sites is nontrivial;
for the odd case no dimer coverings are possible.

\bibitem{cut-points} A cut point is a site that, if removed, would make the graph disconnected.

\bibitem{cut-points-calc} J. O. Fj{\ae}restad (unpublished).

\bibitem{mon-net-zero} Note that the triangle that site
$i$ belonged to (before the bonds connected to $i$ were removed) is replaced by the single bond
connecting the two remaining sites of the original triangle. There is then an arrow constraint on
this bond which is the same as for a linking bond, i.e. both in or both out. Thus the original
triangle constraint has been replaced by a linking bond constraint, giving a net contribution
of 0 to $\Delta \mathcal{N}_c$ from this process.

\bibitem{ndc-kag} A. J. Phares and F. J. Wunderlich, Nuovo Cimento Soc. Ital. Fis. \textbf{101B}, 653 (1988);
V. Elser, Phys. Rev. Lett. \textbf{62}, 2405 (1989).

\bibitem{siddharthan} R. Siddharthan and A. Georges, Phys. Rev. B \textbf{65}, 014417 (2001).

\end{thebibliography}
\end{document}